\documentclass[12pt]{article}

\usepackage[textheight=22.5truecm,textwidth=16.8truecm,top=2.95truecm,left=2.15truecm]{geometry}

\usepackage{
    amsmath,
    amsthm,
    amssymb,
    mathtools,
    ascmac,
    bm,
    tikz,
    braket,
    mathrsfs,
    graphicx,
    subcaption,
    color,
    cite,
    titlesec,
    caption,
    enumitem,
    titling
}
\usepackage{authblk}
\usepackage{hyperref}

\hypersetup{colorlinks,bookmarksopen,bookmarksnumbered,citecolor=black,linkcolor=black,linktocpage,pdfstartview=FitH,urlcolor=black}

\numberwithin{equation}{section}
\numberwithin{figure}{section}

{\makeatletter
\long\def\@makefntext#1{\parindent 1em\noindent 
\@hangfrom{\hbox to 1.8em{\hss$^{\@thefnmark}$}}#1}
\makeatother}

\def\fnum@figure{\textbf{\figurename\nobreakspace\thefigure}}
\def\fnum@table{\textbf{\tablename\nobreakspace\thetable}}
\long\def\@makecaption#1#2{%
  \vskip\abovecaptionskip
  \sbox\@tempboxa{\small #1. #2}%
  \ifdim \wd\@tempboxa >\hsize
    \small #1. #2\par
  \else
    \global \@minipagefalse
    \hb@xt@\hsize{\hfil\box\@tempboxa\hfil}%
  \fi
  \vskip\belowcaptionskip}
\setlist[enumerate]{nosep,leftmargin=16pt}
\setlist[itemize]{nosep,leftmargin=16pt}

\renewcommand{\d}[0]{\mathrm{d}}

\allowdisplaybreaks

\title{
\vspace{-2cm}
\begin{flushright}
\normalsize\textnormal{KUNS-3101, RIKEN-iTHEMS-Report-26} 
\end{flushright}
\vspace{3.5cm} 
Black-hole formation and thermalization\\ in open JT gravity
}

\author[1]{Ryo Adachi \thanks{\href{adachiryo@gauge.scphys.kyoto-u.ac.jp}{\texttt{adachiryo@gauge.scphys.kyoto-u.ac.jp}}}}
\author[2,3]{Rumi Hasegawa\thanks{\href{rumi@muse.sc.niigata-u.ac.jp}{\texttt{rumi@muse.sc.niigata-u.ac.jp}}}}
\author[1]{Takanori Ishii\thanks{\href{mailto:ishiit@gauge.scphys.kyoto-u.ac.jp}{\texttt{ishiit@gauge.scphys.kyoto-u.ac.jp}}}}
\author[3]{Daichi Takeda\thanks{\href{mailto:daichi.takeda@riken.jp}{\texttt{daichi.takeda@riken.jp}}}}

\affil[1]{\small\textit{Theoretical Particle Physics Group
Department of Physics, Kyoto University
Kitashirakawa, Kyoto 606-8502, Japan}
}
\affil[2]{
\textit{Graduate School of Science and Technology, Niigata University, Niigata 950-2181, Japan}
}
\affil[3]{
\textit{iTHEMS, RIKEN, Wako, Saitama 351-0198, Japan}
}

\date{}

\begin{document}
\maketitle

\begin{abstract} 
	Black-hole formation is expected, via holography, to correspond to thermalization in the boundary theory. For open quantum systems, an initial pure state generically evolves into a mixed state irreversibly, suggesting that horizon formation in the bulk should arise. In this paper, we extend the holographic Lindblad prescription to a non-Markovian setting and apply it to JT gravity coupled to a scalar field. Using numerical simulations in the semiclassical and high-temperature regime, we demonstrate the dynamical formation of black holes.
\end{abstract}

\newpage

\tableofcontents

\newpage

\section{Introduction}\label{sec:intro}
According to the holographic principle, a gravitational theory in $(d+1)$ dimensions can be described by an equivalent $d$-dimensional field theory. The most prominent example is the AdS/CFT correspondence \cite{Maldacena:1997re,Gubser:1998bc,Witten:1998qj}, which relates a $(d+1)$-dimensional AdS spacetime to a $d$-dimensional CFT. Although a complete theory of quantum gravity is still lacking, AdS/CFT makes it possible to study black hole dynamics in terms of better-understood field theories.

In AdS/CFT, it is well known that a stationary black hole corresponds to a thermal equilibrium state of the boundary theory \cite{Witten:1998zw}.
It is therefore natural to interpret black-hole formation in the bulk as thermalization in the boundary theory.
This viewpoint has been explored in models of gravitational collapse, where black-hole formation and subsequent relaxation in the bulk encode thermalization or isotropization in the boundary theory; see, for example, \cite{Danielsson:1999fa,Chesler:2008hg,Balasubramanian:2011ur} and subsequent works.
These processes describe thermalization in isolated systems: the final states are not literally thermal mixed states, but rather pure states for which expectation values of a suitable class of observables appear thermal.

On the other hand, when a boundary system is coupled to an external heat bath, an initially prepared pure state generally evolves into a mixed state.
In other words, purity is generally lost due to a transfer of information to environmental degrees of freedom. If the reduced boundary theory admits a semiclassical bulk description, this inaccessible information
should have a geometric representation. A natural possibility is that it is encoded behind a bulk horizon. We therefore expect horizon formation to be a generic feature of open AdS/CFT.
\footnote{
It is natural that no horizon forms when a low-energy state evolves in contact with a low-temperature bath. However, horizon formation is still possible even if the bath is cold, provided that gravitational collapse occurs before thermalization.
}
From this viewpoint, the entropy production and irreversibility of the reduced dynamics would be interpreted as the growth of an inaccessible bulk region bounded by a horizon.

In this paper, we investigate black-hole formation in the holographic description of open systems.
We first extend the holographic prescription for Lindblad dynamics proposed in \cite{Ishii:2025qpy} to a non-Markovian setting, where the environment is a Gaussian bath initially prepared in a thermal Gibbs state.
We then apply this framework to the theory consisting of Jackiw-Teitelboim (JT) gravity\cite{Jackiw:1984je,Teitelboim:1983ux} and a scalar field \cite{Almheiri:2014cka,Engelsoy:2016xyb,Maldacena:2016upp}, where the scalar field is dual to the boundary operator coupled to the bath. JT gravity is a $(1+1)$-dimensional theory of gravity in which the dilaton is coupled to the metric. In our model, the scalar field couples only to the metric minimally and does not interact directly with the dilaton. In this case, the dilaton can be integrated out, and the gravitational sector reduces to the Schwarzian action of the boundary embedding function $t(u)$ \cite{Maldacena:2016upp,Engelsoy:2016xyb}. As a first step, we consider the regime where the bath is at high temperature and the bulk is semiclassical, and analytically derive the equation of motion for the Schwarzian theory, yielding an integro-differential equation, \textit{i.e.}, non-Markovian dynamics.

By numerically solving the equation of motion, we find that $t(u)$ approaches a constant exponentially, at late times. This behavior is consistent with the ordinary AdS$_2$ black hole, thereby supporting our expectation. 
We also find that horizon formation can occur when the boundary operator dual to the scalar field is driven by an external source rather than coupled to a bath. This may correspond to thermalization in isolated systems, and should therefore be distinguished from open AdS/CFT.

A closely related setup was considered in \cite{Gaikwad:2022jar}, where the Liouville mode \(\log \dot t(u)\) is directly coupled to external scalar fields defined on a flat half-space, and the resulting dynamics was used to study black-hole evaporation numerically.
A related influence-functional approach to Renyi and entanglement entropies in open holographic systems, including matter-coupled JT gravity, was discussed in \cite{Pelliconi:2023ojb}.
By contrast, in the present work, our starting point is a general open AdS/CFT dictionary with a Gaussian bath, which we apply to the JT gravity model described above.
In our construction, the bath couples to the boundary theory through an operator \(O\), and we therefore introduce a bulk scalar field dual to \(O\).
Furthermore, our focus is not black-hole evaporation but black-hole formation.

The real-time AdS/CFT prescription based on the Schwinger--Keldysh formalism, which we use in this work, was developed in
\cite{Skenderis:2008dh,Skenderis:2008dg,vanRees:2009rw}
(see also \cite{Glorioso:2018mmw} for another proposal).
Open AdS/CFT dictionaries based on modified boundary conditions or double-trace couplings have also been developed in
\cite{Aharony:2006hz,Karch:2023wui,Geng:2023ynk,Karch:2025hof}.
Our prescription differs from these approaches in that it directly describes the reduced dynamics of the target system after integrating out the environment.
Related holographic descriptions of open quantum systems and stochastic dynamics include
\cite{Son:2009vu,deBoer:2008gu,Jana:2020vyx}, where a quark or another field-theoretic sector is coupled to a holographic environment.

The organization of this paper is as follows. In section \ref{sec:Open JT}, we first explain the prescription for treating boundary theories coupled to a heat bath within AdS/CFT. We then apply it to JT gravity coupled to a scalar field and derive the corresponding equation of motion. In section \ref{sec:Numerical verification}, we numerically analyze the equation of motion obtained in section \ref{sec:Open JT} and show that the asymptotic behavior of the solution coincides with that of a stationary black hole solution in JT gravity. Section \ref{sec:Summery and discussion}  is devoted to conclusions and discussion. Details of calculations omitted in the main text are presented in the Appendix.

\section{Semiclassical open JT gravity with scalar field}\label{sec:Open JT}

In this section, we first describe the gravitational picture in general AdS/CFT when the boundary theory is coupled to a heat bath. We assume that the bath is initially in a Gibbs state and that its dynamics is Gaussian, so that the influence functional can be approximated up to quadratic order. We then apply this prescription to JT gravity coupled to a scalar field and derive the semiclassical equation of motion.

\subsection{AdS/CFT with non-Markovian bath}
We take the action of the total system consisting of the CFT and the heat bath to be 
\begin{align} I_{\mathrm{tot}} = I[\varphi] + I_{B}[\psi]+\int \mathrm{d}^d x\, g(x)O(x)\psi(x) . 
\end{align} 
Here we use the shorthand notation $x=(t,\vec{x})$. The field $\varphi$ collectively denotes the degrees of freedom of the CFT, and $I[\varphi]$ is the CFT action. The operator $O$ is a composite operator of $\varphi$ and is taken to be a scalar primary. The term $I_B[\psi]$ denotes the action of the heat bath, where $\psi$ is its fundamental scalar field. We take $g(x)$ to be a general coordinate-dependent function. As the initial condition, we assume that the CFT is in the vacuum state, while the heat bath is in the Gibbs state, $e^{-\beta H_B}/\mathrm{Tr}\,e^{-\beta H_B}$.

\begin{figure}
    \centering
    \includegraphics[width=0.7\linewidth]{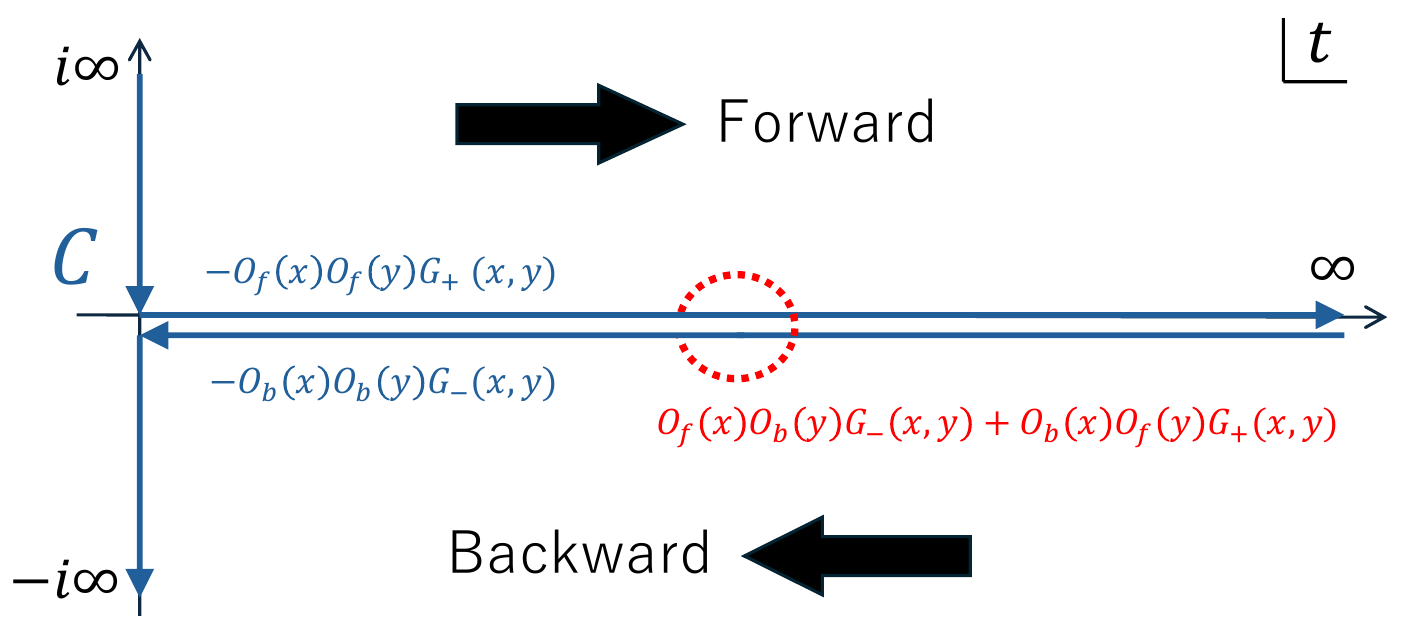}
    \caption{The Schwinger--Keldysh contour $C$ on the boundary. 
    The evolution along the imaginary-time direction prepares the initial state.
    Here, we have defined
    $G_\pm(x,y) \coloneq g(x)g(y)\left(\pm i\eta(x-y)\Theta(x^0-y^0) + \nu(x-y)\right)/2.$}
    \label{fig_bdy_contour}
\end{figure}

Since the heat bath is assumed to be Gaussian, it can be integrated out \cite{FEYNMAN1963118, CaldeiraLeggett}, yielding the partition function (see \cite{Liu:2018kfw} for review):
\footnote{
Since $Z$ contains no source term, it is not a generating functional but simply the trace of the state, and therefore evaluates to a trivial constant. In this paper, our main interest is to determine the bulk dual associated with the saddle point of $Z$. Accordingly, we retain this formal path-integral expression without introducing sources.
}
\begin{align} Z& = \int \mathcal{D}\varphi\;e^{iI[C;\varphi]}\exp\left[ -\int \mathrm{d}^d x  \mathrm{d}^d y\Bigg\{g(x)g(y)\right.\notag\\&\left.\times\left(2iO_a(x)\eta(x-y)\Theta(x^0-y^0)O_r(y) + \frac{1}{2}O_a(x)\nu(x-y)O_a(y)\right)\Bigg\} \right] ,\label{genfncft} 
\end{align} 
where $\Theta$ is the step function, and $C$ denotes the time contour along which the path integral is performed (Fig.\ \ref{fig_bdy_contour}). The path integral over the Euclidean segment prepares the vacuum state of the CFT. We also define 
\begin{align}
O_a \coloneq O_f -O_b,\qquad O_r \coloneq \frac{O_f+O_b}{2},
\end{align}
where the subscripts $f$ and $b$ denote the forward and backward branches, respectively.

The functions $\Theta(x^0)\eta(x)$ and $\nu(x)$ are, respectively, the retarded Green's function and the symmetrized two-point function of the heat bath. They play the roles of the dissipation and noise kernels for the target system. By the Kubo--Martin--Schwinger condition, they are related as 
\begin{gather} \label{eq: FTR}
\tilde{\nu} (k) = -\coth \frac{k^0\beta}{2} \mathrm{Im} \,\tilde{\eta}(k). 
\end{gather} 
Here, $\tilde{\nu}$ and $\tilde{\eta}$ denote the Fourier components of $\nu$ and $\eta$, respectively, and \eqref{eq: FTR} is known as the fluctuation--dissipation relation. This relation implies that, in the high-temperature regime, $\nu$ dominates over $\eta$. In what follows, we therefore consider the high-temperature limit and neglect the contribution of $\eta$. We also define $\mathcal{N}(x,y) \coloneq g(x)g(y)\nu(x-y)$. Since $\nu$ is an even function, we have $\mathcal{N}(x,y) = \mathcal{N}(y,x)$.

\begin{figure}
    \centering
    \includegraphics[width=0.5\linewidth]{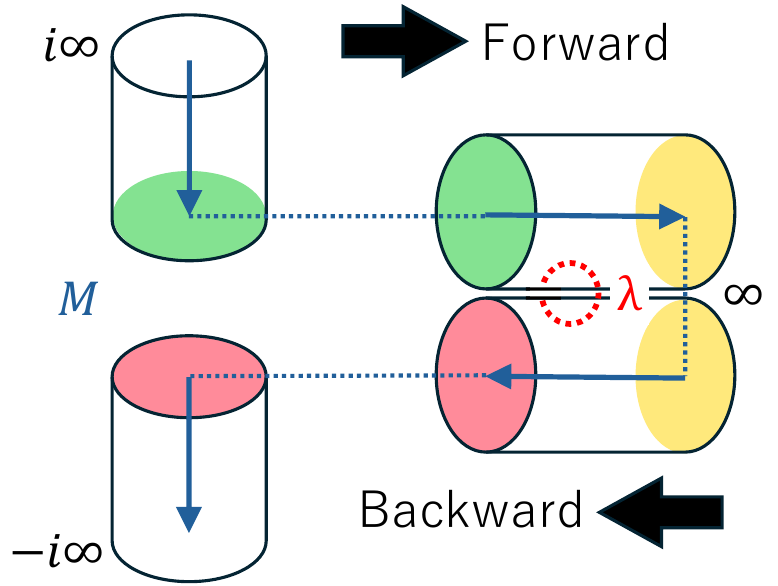}
    \caption{The bulk spacetime $M$ corresponding to the contour $C$ in Fig.\ \ref{fig_bdy_contour}. Each field is glued ``smoothly'' across the matching surfaces indicated by the same color. The quantity $\lambda$ represents the colored noise injected through the boundary condition of the scalar field.
    \label{fig_bulk_contour}}
\end{figure}

Applying the GKPW relation \cite{Gubser:1998bc,Witten:1998qj} to \eqref{genfncft}, we obtain the following expression in the high-temperature limit of the heat bath:
\begin{gather} 
Z = \int \mathbb{D}\Phi\mathcal{D}\lambda \;e^{iS[M;\Phi,\lambda]}\times\exp\left[ -\frac{1}{2}\int \mathrm{d}^d x \mathrm{d}^d y\; \lambda(x)\hat{\mathcal{N}}(x,y)\lambda(y)\right] .\label{genfnads}
\end{gather} 
The derivation of \eqref{genfnads} from \eqref{genfncft} is explained in Appendix \ref{app:bath AdS/CFT}, where we treat the more general case retaining $\eta$. Here, $\mathbb{D}$ denotes the measure for the bulk path integral, $\Phi$ is the scalar field dual to $O$, $S$ is the bulk action, and $\hat{\mathcal{N}}$ is the inverse of $\mathcal{N}$ with respect to convolution.\footnote{
The inverse $\hat{\mathcal{N}}$ is defined only on the subspace where the coupling $g(x)$ is supported. Out of the support, we understand $\lambda(x) \equiv 0$.
} The manifold $M$ is the spacetime on which the path integral is performed (Fig.\ \ref{fig_bulk_contour}), namely, an asymptotically AdS spacetime whose asymptotic boundary is given by $C\times (\mathrm{space})$. 
The quantity $\lambda$ appearing as an argument of the bulk action specifies the boundary condition for the scalar field, which is
\begin{gather} 
\Phi_{f,b}(z,x)\sim z^{d-\Delta}\lambda(x),\qquad \Phi_E(z,x) \sim \mathcal{O}(z^\Delta)\qquad(z\sim0),
\label{bdy_cond_scaler}
\end{gather}
in the Fefferman--Graham coordinates:
\begin{align} 
\mathrm{d}s^2 \sim \frac{\pm\mathrm{d}t^2 + \mathrm{d}\vec{x}^2+\mathrm{d}z^2}{z^2} \qquad (z\sim0).
\end{align}
Here, $\Delta$ is the scaling dimension of $O$, and $\Phi_E$ denotes the scalar field on the Euclidean segment. In addition, to ensure that the variational principle is well-defined, we impose the condition that each field be smooth across the gluing surfaces connecting the different parts of $M$ \cite{Skenderis:2008dh,Skenderis:2008dg}. Equation \eqref{genfnads} is a generalization of the prescription of \cite{Ishii:2025qpy}, in which the incoming white noise is replaced by colored noise.

In \cite{Ishii:2025qpy}, the bulk degrees of freedom $\Phi$ were first integrated out, which is done by evaluating the on-shell action in the classical regime, and the integration over $\lambda$ was performed afterward. This then yields a classical bulk solution for each history of $\lambda$. Each such bulk solution simply corresponds to unitary time evolution in the presence of an external source $\lambda$. In the present paper, we instead first integrate out $\lambda$ together with the bulk scalar field whose boundary value is $\lambda$, so that the gravitational degrees of freedom incorporate the openness of the boundary.

\subsection{Application to JT gravity with matter}
We now apply \eqref{genfnads} to JT gravity coupled to a scalar field. On the Lorentzian segments, the bulk action is given by
\begin{align}
&S[g,\phi,\Phi] = S_g[g,\phi] + S_m[g,\Phi] + S_{ct}[g,\Phi],\\
&S_g \coloneq -\frac{1}{16\pi G}\left[
\int_M \mathrm{d}^2 x \sqrt{-g}\phi (R+2)
+ 2\int_{\partial M} \mathrm{d}u \sqrt{-h}\phi(K-1)
+ 2 \sum_j \phi\theta_j
\right],
\label{action_JT}\\
&S_m \coloneq -\frac{1}{2}\int_{M} \mathrm{d}^2 x \sqrt{-g}(\partial_\mu \Phi \partial^\mu \Phi + m^2\Phi^2),
\label{action_scaler}
\end{align}
and similarly on the Euclidean segments. Here $u$ denotes the time coordinate of the boundary theory, the AdS radius is set to $1$, and $m^2 = \Delta(\Delta-1)$. The boundary conditions for the metric and the dilaton $\phi$ are taken to be
\begin{gather}
\mathrm{d}s^2|_{\partial M} = \pm\frac{\mathrm{d}u^2}{\epsilon^2},\qquad\phi|_{\partial M} = \frac{1}{\epsilon}
\label{bdy_cond_JT}
\end{gather}
where $\epsilon$ is a cutoff parameter introduced to define the boundary conditions at the asymptotic boundary, and we assign $+$ ($-$) to the Euclidean (Lorentzian) segments. The third term in $S_g$ is the Hayward term \cite{Hayward:1993my}, which is required in order for the variational principle to be well-defined when the boundary of the manifold is not smooth. The quantity $\theta_j$ in the Hayward term is the (boost) angle between the normal vector to the asymptotic boundary and that to the gluing surface between different segments, with $j$ labeling the corners on the boundary.\footnote{For example, on a Euclidean segment, if $n_a$ and $n_g$ denote the normal vectors to the asymptotic boundary and the geodesic boundary, respectively, then $\theta_j = \sin^{-1}(n_a\cdot n_g)$.}
The term $S_{ct}$ is a counterterm introduced to cancel the divergence of $S_m$ near the asymptotic boundary, and for $1/2<\Delta<3/2$ it is given by \cite{deHaro:2000vlm}
\begin{gather}
S_{ct} = \frac{\Delta-1}{2}\int_{\partial M} \mathrm{d}u \sqrt{-h}\Phi^2.
\end{gather}
 We also impose the boundary condition \eqref{bdy_cond_scaler} on the scalar field, taking care regarding the relation between the Poincar\'e coordinates and the boundary time $u$ (see Appendix \ref{app:On-shell action}).

In \eqref{action_JT} the dilaton $\phi$ acts as a Lagrange multiplier, and integrating out $\phi$ constrains the bulk metric to be locally AdS$_2$. Therefore, the only remaining gravitational degree of freedom is the embedding of the boundary together with the topology of the manifold. In what follows, we further restrict attention to the semiclassical regime on the disk topology, where the only remaining degree of freedom is the boundary embedding $(t(u),z(u))$.
Although global AdS$_2$ is two-sided, we focus only on one side and use Poincar\'{e} coordinate.
Under the boundary condition \eqref{bdy_cond_JT}, the gravitational action then reduces to the Schwarzian action \cite{Maldacena:2016upp,Engelsoy:2016xyb}:
\begin{align}
&S_\mathrm{Sch} [t]\coloneq -\frac{1}{8\pi G}\int \mathrm{d}u\, 
\{t,u\}
\label{action_Schwarzian}\\
&\{t,u\} \coloneq \frac{\dddot{t}}{\dot{t}}-\frac{3}{2}\left(\frac{\ddot{t}}{\dot{t}}\right)^2,\qquad \dot{t}\coloneq\frac{\mathrm{d}}{\mathrm{d}u}t.
\end{align}

To study the gravitational dynamics, we integrate out the dilaton and the bulk scalar field. The partition function then becomes
\begin{align}
&Z =
\int \mathcal{D}t \,e^{iS_\mathrm{Sch}[t_f]-iS_\mathrm{Sch}[t_b] + iW[t_f,t_b]},\\
&e^{iW[t_f,t_b]} \coloneq
\int\mathcal{D}\lambda\;e^{iS_{m,\mathrm{eff}}[t_f,t_b;\lambda]}
\times \exp\left[-\frac{1}{2}
\int_0^\infty \mathrm{d}u_1 \int_0^\infty \mathrm{d}u_2\,
\lambda(u_1)\hat{\mathcal{N}}(u_1,u_2)\lambda(u_2)
\right],\\
&iS_{m,\mathrm{eff}} [t_f,t_b;\lambda]
\coloneq
-\left( \Delta - 1/2\right)C_\Delta
\int_0^\infty \mathrm{d}u_1 \,\mathrm{Pf}\int_0^\infty \mathrm{d}u_2 \,\lambda(u_1)\lambda(u_2)\notag\\
&\times \left[
e^{-i\pi\Delta}
\frac{(\dot t_f(u_1)\dot t_f(u_2))^\Delta}{|t_f(u_1)-t_f(u_2)|^{2\Delta}}
\right.\notag\\
&\left.\qquad
-2e^{i\pi\Delta \text{sgn}(t_f(u_1)-t_b(u_2))}
\frac{(\dot t_f(u_1)\dot t_b(u_2))^\Delta}{|t_f(u_1)-t_b(u_2) |^{2\Delta}}
+ e^{i\pi\Delta}
\frac{(\dot t_b(u_1)\dot t_b(u_2))^\Delta}{|t_b(u_1)-t_b(u_2) |^{2\Delta}}
\right],
\label{on-shellaction}\\
&C_\Delta \coloneq \Gamma(\Delta)/(\sqrt{\pi}\Gamma(\Delta-1/2)).
\end{align}
Here, $\mathrm{Pf}$ denotes the Hadamard regularization of the integral. Since we are working in the high-temperature and semiclassical limit, the factor $\det(\nabla^2 - m^2)$ arising from fluctuations around the saddle point of the bulk scalar field is relatively small compared with the terms shown above, and therefore we have neglected it.
The derivation of \eqref{on-shellaction} is shown in Appendix \ref{app:On-shell action}.

The equation of motion is obtained from the saddle-point condition of the effective action,
\begin{align}
    S[t_f,t_b] = S_\mathrm{Sch}[t_f]-S_\mathrm{Sch}[t_b] + W[t_f,t_b].
\end{align}
Assuming that the saddle point lies at $t_f(u) = t_b(u)$, one may first vary with respect to $t_f$ and then set $t_f(u) = t_b(u) = t(u)$.\footnote{
In the full Schwinger--Keldysh theory, the difference field
\(t_a:=t_f-t_b\), sometimes called the quantum field, encodes fluctuations
around the classical saddle and becomes relevant beyond the semiclassical saddle-point approximation considered here.
} In this case, the $\lambda$ integral contained in $W$ can be carried out explicitly after the variation (see Appendix \ref{app:eom}). One then obtains the following equation of motion:
\begin{align}
&\frac{1}{8\pi G}\frac{1}{\dot{t}}\frac{\mathrm{d}}{\mathrm{d}u}
\{t,u\} = (4\Delta-2)\sin(\pi\Delta)C_\Delta\mathrm{Pf}\int_0^u \mathrm{d}u' \,\mathcal{A}(u,u'),
\label{eom}\\
&\mathcal{A}(u,u')\coloneq
\frac{2\Delta(\dot{t}(u)\dot{t}(u'))^\Delta}{|t(u)-t(u')|^{2\Delta+1}}\mathcal{N}(u,u')
+ \Delta\frac{\partial}{\partial u}\left(\frac{\dot{t}(u)^{\Delta-1}\dot{t}(u')^\Delta}{|t(u)-t(u')|^{2\Delta}}\mathcal{N}(u,u') \right)
\end{align}
In \eqref{eom}, the left-hand side originates from the gravitational action, while the right-hand side represents the contribution of the scalar field and $\lambda$.

\section{Numerical verification of black-hole formation}\label{sec:Numerical verification}
In this section, we numerically solve the equation of motion \eqref{eom} in order to test our expectation that, when the boundary system is open, black-hole formation occurs in the dual bulk. 

\subsection{Black hole solutions in pure JT gravity}\label{subsec:pure JT}
Before presenting the numerical results, let us review the criterion for black-hole formation in JT gravity. To this end, we briefly review the black hole solutions of pure JT gravity without matter fields.

In the absence of matter fields, the equation of motion derived from the Schwarzian action \eqref{action_Schwarzian} is given as
\begin{equation}
\frac{\mathrm{d}}{\mathrm{d}u} \{t,u\} = 0 \implies \{t,u\} = -E,
\end{equation}
where the constant $E$ is interpreted as the energy of the system \cite{Maldacena:2016upp,Engelsoy:2016xyb}. Up to PSL($2,\mathbb{R}$) transformations\footnote{This is the symmetry given by $t \to (a t+b)/(c t+d)$ with $ad-bc=1$.}, the solution of this equation is written as
\begin{equation}
t(u) = \frac{\beta}{\pi}\tanh\left(\frac{\pi}{\beta}u\right),\qquad E = \frac{2\pi^2}{\beta^2}.
\label{BHsolution}
\end{equation}
The parameter $\beta$ can be interpreted as the inverse temperature of the black hole, since it determines the periodicity of \eqref{BHsolution} after analytically continuing $u$ to Euclidean time.

\begin{figure}
    \centering
    \includegraphics[width=0.4\linewidth]{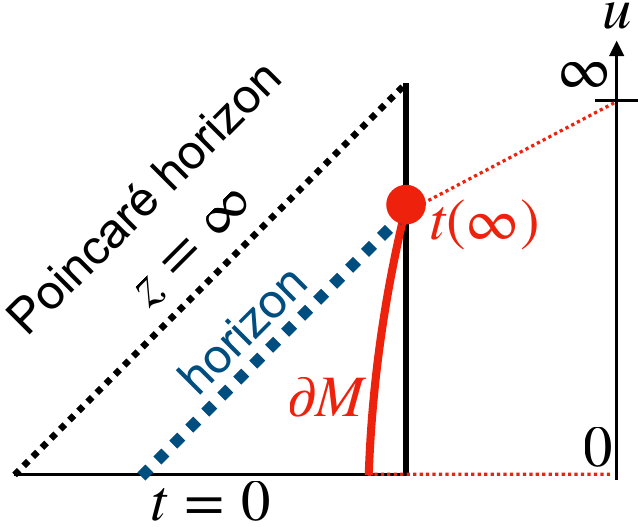}
    \caption{Dynamical formation of AdS$_2$ black hole.}
    \label{fig:horizon_in_Poincare}
\end{figure}

This solution behaves at late times as
\begin{align}
t(u) \sim \frac{\beta}{\pi} \left(1-2e^{-2\pi u/\beta}\right)\qquad(u\to \infty).
\end{align}
The fact that $t(u)$ approaches a finite value means that the boundary is embedded at a finite Poincaré time, implying the existence of a region in the bulk that cannot communicate causally with the boundary, as in Fig.\ \ref{fig:horizon_in_Poincare}. In other words, the presence or absence of black-hole formation can be diagnosed from the late-time behavior of $t(u)$ as a function of the boundary time $u$, namely by whether $t(u)$ approaches a constant. In particular, for the solution \eqref{BHsolution}, $\log\dot{t}(u)$ becomes linear in $u$ at large $u$, with slope $-2\pi/\beta$. Therefore, if a numerical plot of $\log\dot{t}(u)$ exhibits linear behavior at late times, one may conclude that the system thermalizes to a stationary black hole with a temperature determined by the corresponding slope.

\subsection{Numerical solutions in open JT gravity}
Now we numerically solve the equation of motion and examine whether black-hole formation occurs. 
We set $8\pi G = 1$ and $\Delta = 3/5$.
Also, we ignore the factor $(4\Delta-2)\sin(\pi\Delta)C_\Delta$ in \eqref{eom}, by absorbing this factor into the normalization of the coupling $g(u)$.
The last condition is nothing but $\tau(u) = u$, since no matter excitation is introduced on the Euclidean segment. Reflecting this, we impose the following initial conditions on the Lorentzian segment:
\begin{equation}
t(0) = 0,\quad \dot{t}(0) = 1,\quad\ddot{t}(0) = 0, \quad\dddot{t}(0) = 0
\end{equation}
In the calculations below, we take the colored noise to be
\begin{equation}
\nu(u) = \frac{\epsilon}{\pi}\frac{1}{\epsilon^2+u^2}
\label{normalized_delta}
\end{equation}
This choice of $\nu$ is obtained by taking the heat bath to be a Feynman--Vernon bath, choosing a regularized Ohmic spectrum, and then taking the high-temperature limit (Appendix \ref{app:bath AdS/CFT}).
\begin{figure}[t]
    \begin{tabular}{cc}
      \begin{minipage}[t]{0.45\hsize}
        \centering
        \subcaption{$t(u)$}
        \includegraphics[keepaspectratio, scale=0.35]{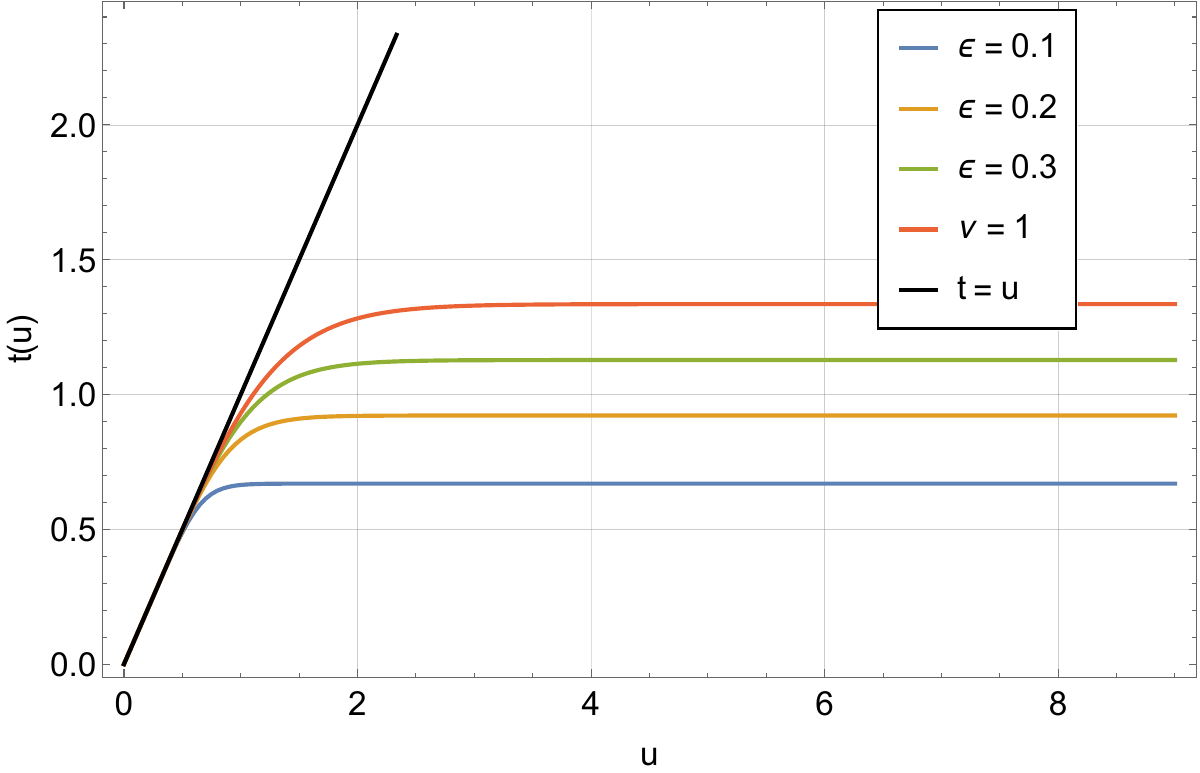}
        \label{fig_t_cubic_con}
      \end{minipage} &
      \begin{minipage}[t]{0.45\hsize}
        \centering
        \subcaption{$\dot{t}(u)$}
        \includegraphics[keepaspectratio, scale=0.35]{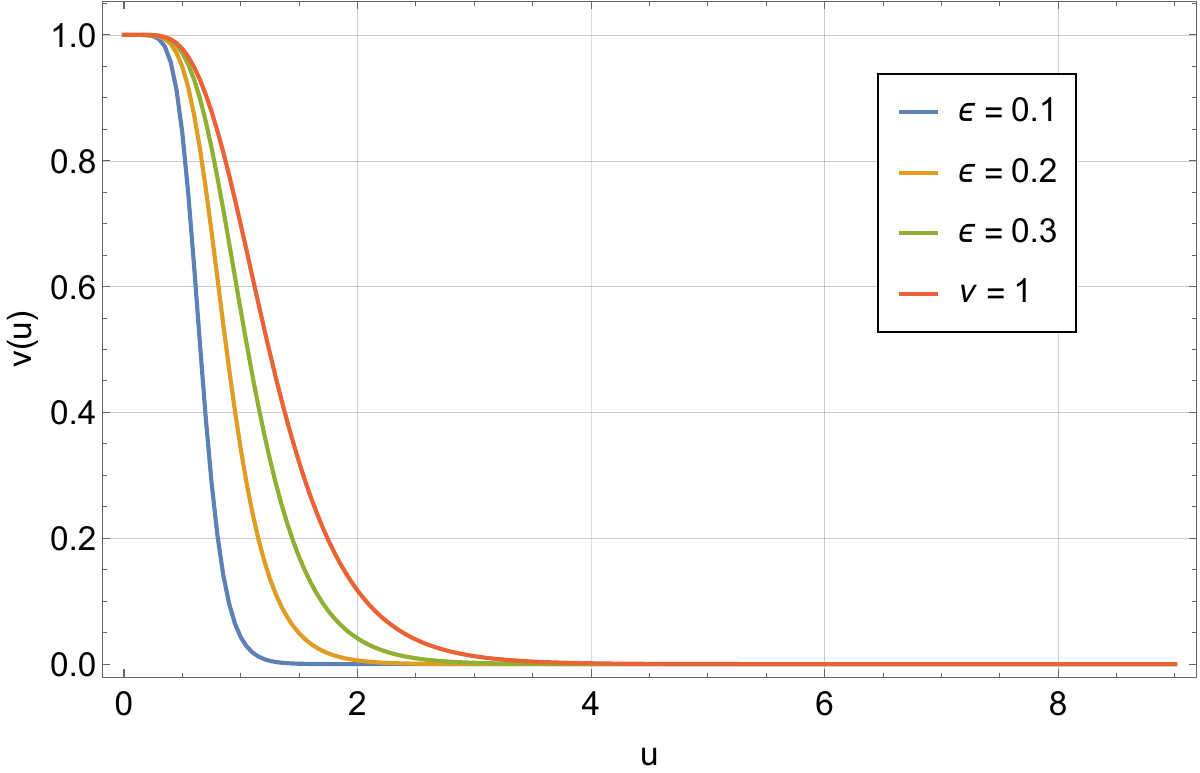}
        \label{fig_tdot_cubic_con}
      \end{minipage} \\[40pt]
   
      \begin{minipage}[t]{0.45\hsize}
        \centering
        \subcaption{$\log(\dot{t}(u))$}
        \includegraphics[keepaspectratio, scale=0.35]{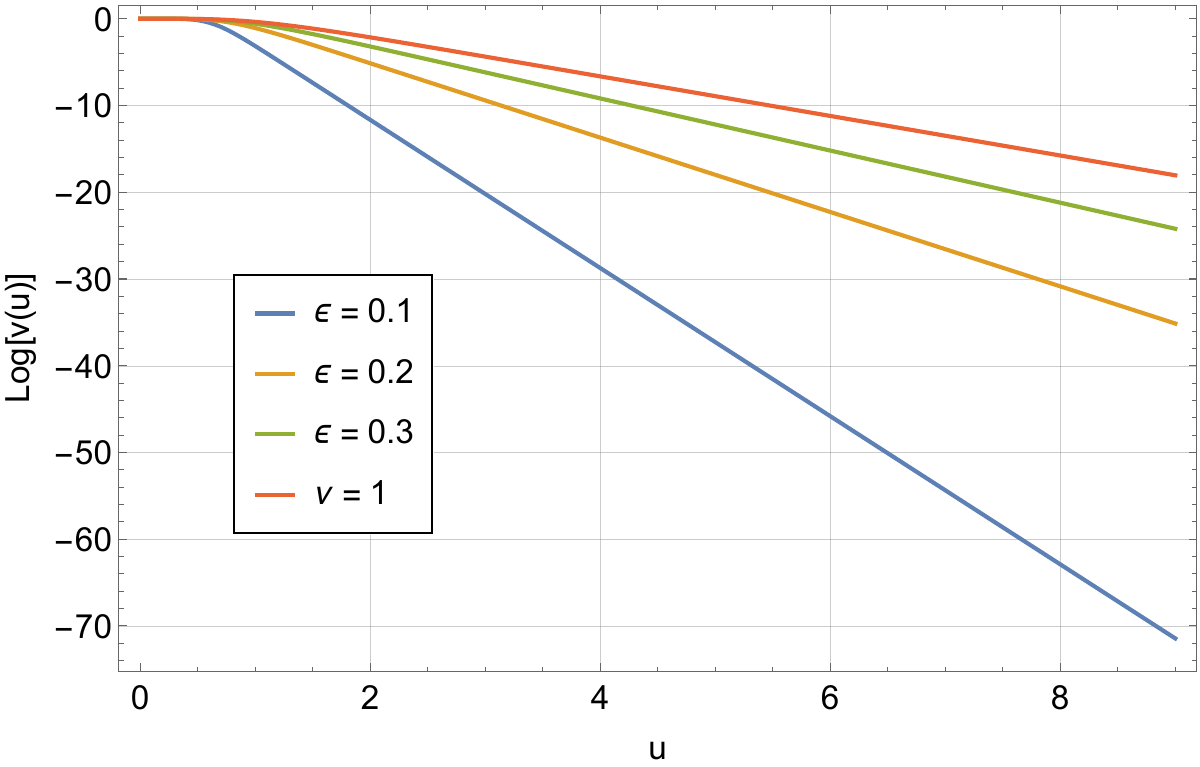}
        \label{fig_logtdot_cubic_con}
      \end{minipage} &
      \begin{minipage}[t]{0.45\hsize}
        \centering
        \subcaption{$\{t,u\}$}
        \includegraphics[keepaspectratio, scale=0.35]{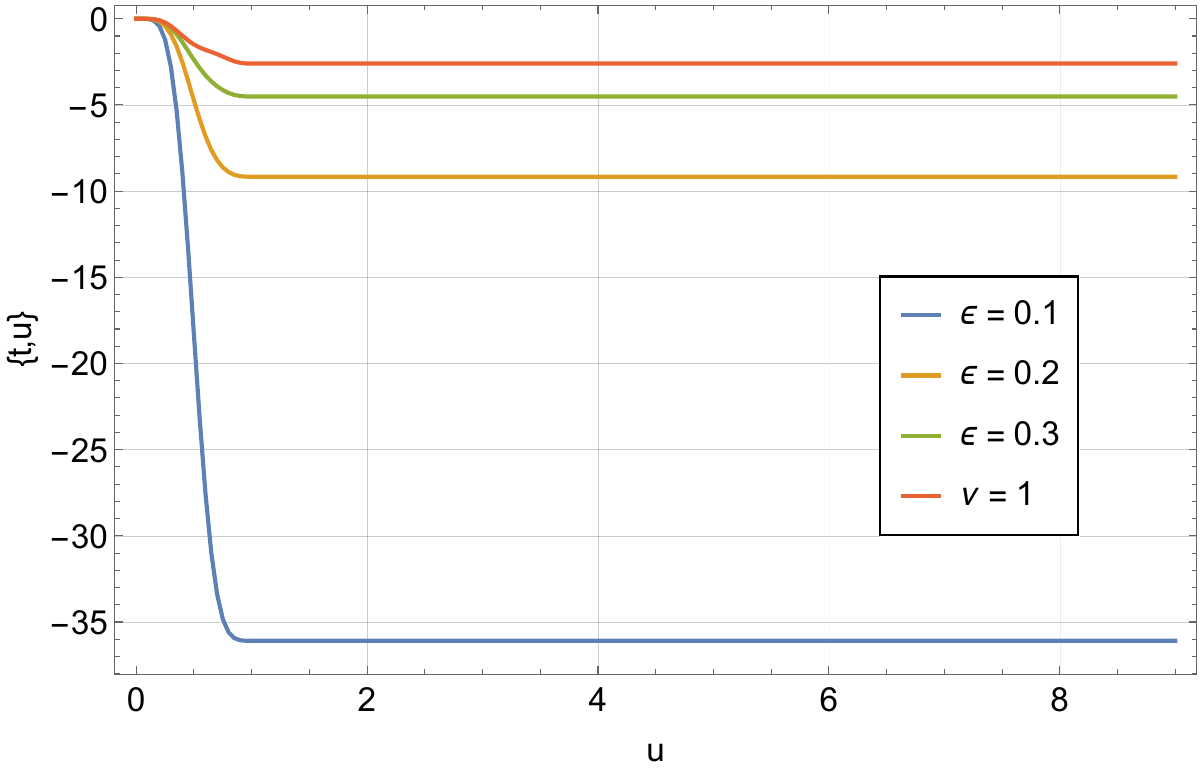}
        \label{fig_Sch_cubic_con}
      \end{minipage}
    \end{tabular}
    \caption{Numerical results for \eqref{g_cubic}. The blue, yellow, and green curves correspond to the results for $\epsilon = 0.1,\,0.2,\,0.3$, respectively. The red curve shows the result for $\nu=1$, which corresponds to the isolated system.
    }
    \label{fig_cubic_con}
  \end{figure}
In addition to varying \(\epsilon\) in the numerical analysis, we also consider, as a limiting case, the choice \(\nu(u)=1\).
As can be seen from the derivation in Appendix \ref{app:eom}, this choice reproduces the semiclassical equation for an isolated boundary theory driven by the external source term \(\int du\,g(u)O(u)\), rather than coupled to a heat bath.

Fig.\ \ref{fig_cubic_con} shows the results for
\begin{align}
g(u) &= \begin{cases}
2^6u^3(1-u)^3&\text{$u \in [0,1]$}\\
0\;&\text{$u \notin[0,1]$}
\end{cases}
\label{g_cubic}
\end{align}
This corresponds to the case in which the boundary theory is smoothly coupled to the heat bath at time $u=0$ and decoupled again at time $u=1$.
In Fig.\ \ref{fig_cubic_con}, $t(u)$  approaches a finite value as $u\to\infty$. As explained in Section \ref{subsec:pure JT}, this implies that an event horizon forms in the bulk, in agreement with our expectation. Furthermore, the plot of $\log \dot{t}(u)$ shows linear behavior at large $u$, indicating that the stationary black hole of JT gravity \eqref{BHsolution} is realized. One can also see that smaller values of $\epsilon$ correspond to noise closer to white noise, leading to more rapid thermalization into a higher-temperature black hole.
In Fig.\ \ref{fig_cubic_con}, the red curve ($\nu=1$) corresponds to the case in which the boundary theory is an isolated system under an external field $g(u)$. Even in this case, black-hole formation is observed, similarly to the other cases. Our expectation was only that time evolution from a pure state to a mixed state leads to black-hole formation in the bulk, and therefore does not impose any restriction on the dynamics of isolated systems. Hence, this result does not conflict with our expectations.
\begin{figure}[t]
    \begin{tabular}{cc}
      \begin{minipage}[t]{0.45\hsize}
        \centering
        \subcaption{$t(u)$}
        \includegraphics[keepaspectratio, scale=0.35]{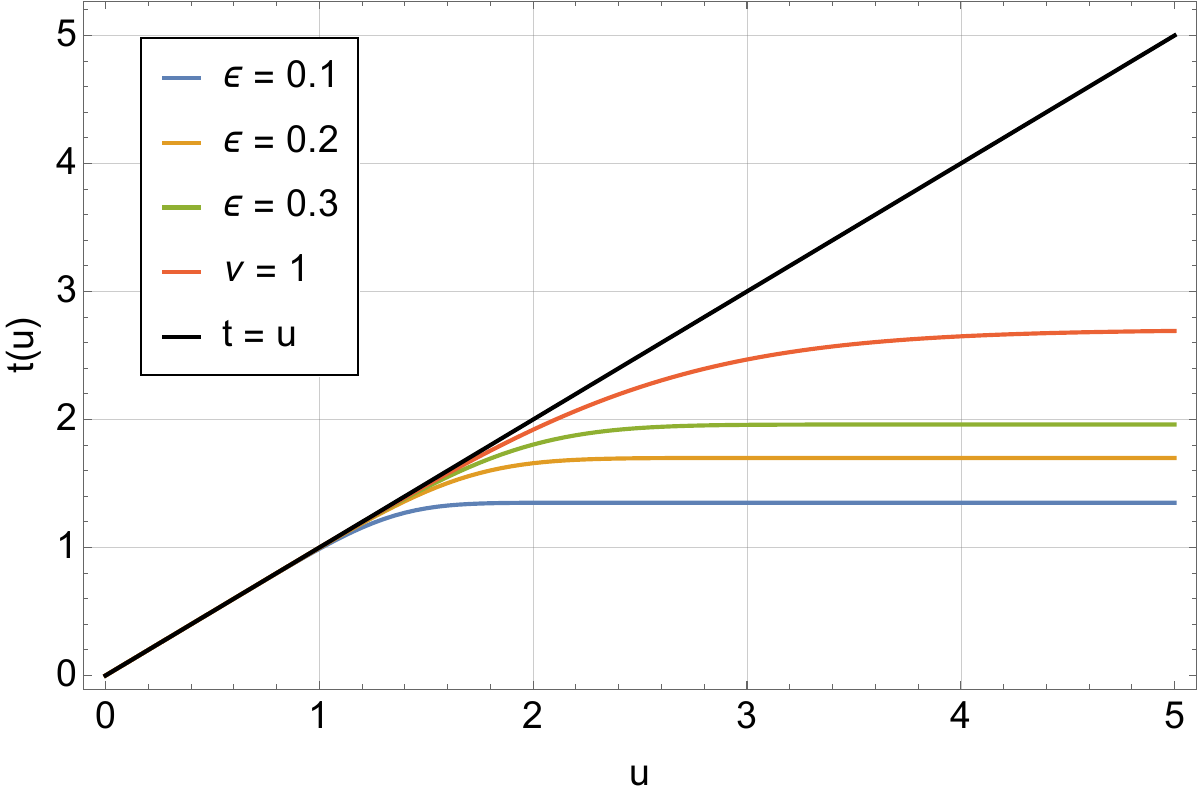}
        \label{fig_t_tanh}
      \end{minipage} &
      \begin{minipage}[t]{0.45\hsize}
        \centering
        \subcaption{$\dot{t}(u)$}
        \includegraphics[keepaspectratio, scale=0.35]{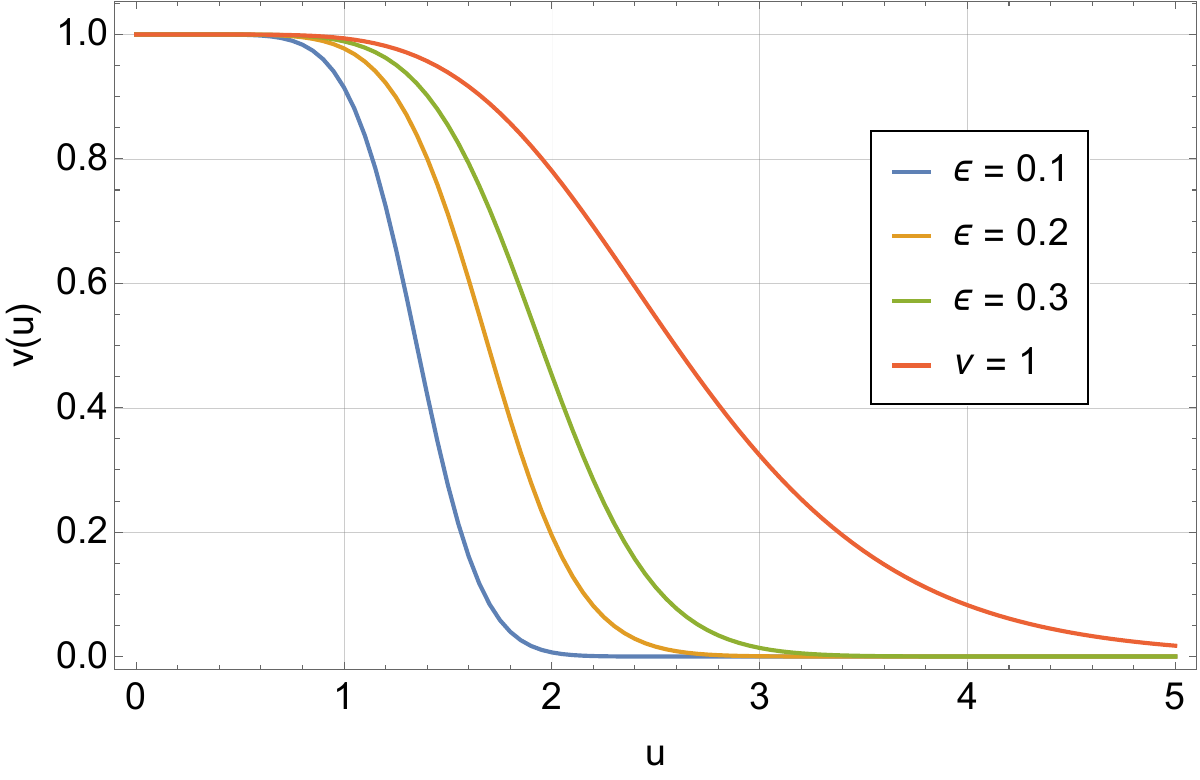}
        \label{fig_tdot_tanh}
      \end{minipage} \\[40pt]
   
      \begin{minipage}[t]{0.45\hsize}
        \centering
        \subcaption{$\log(\dot{t}(u))$}
        \includegraphics[keepaspectratio, scale=0.35]{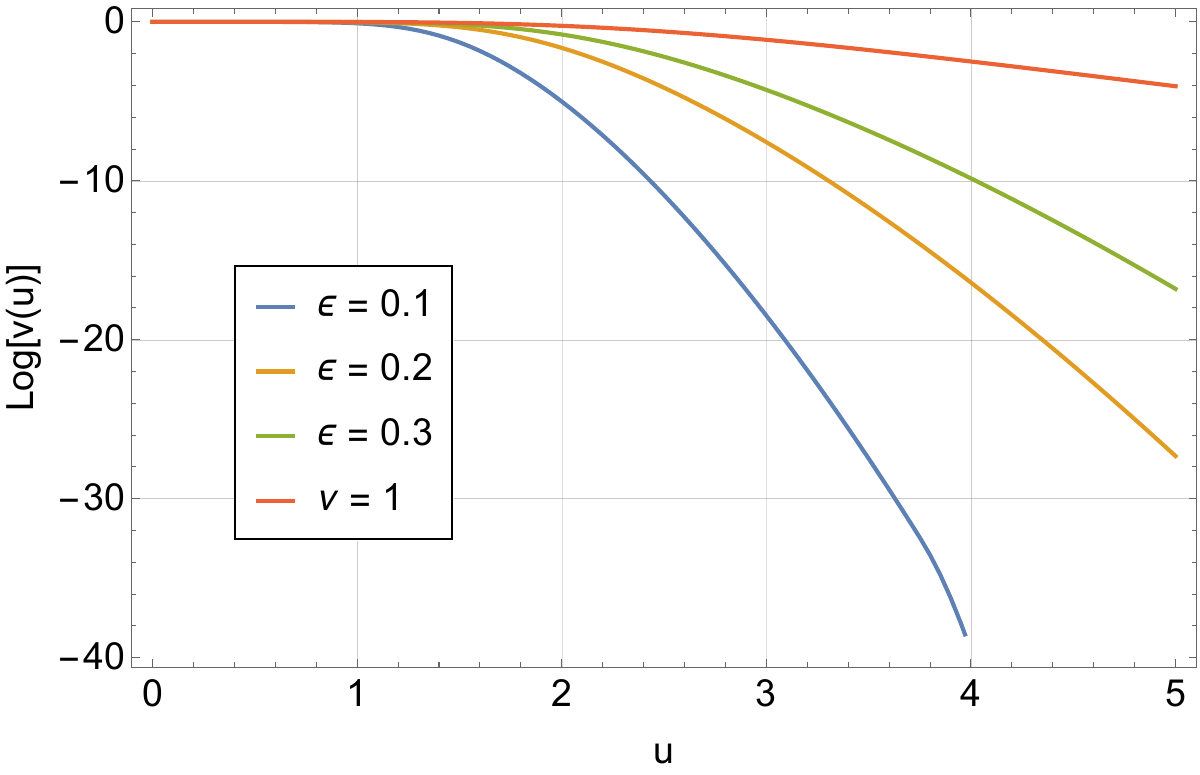}
        \label{fig_logtdot_tanh}
      \end{minipage} &
      \begin{minipage}[t]{0.45\hsize}
        \centering
        \subcaption{$\{t,u\}$}
        \includegraphics[keepaspectratio, scale=0.35]{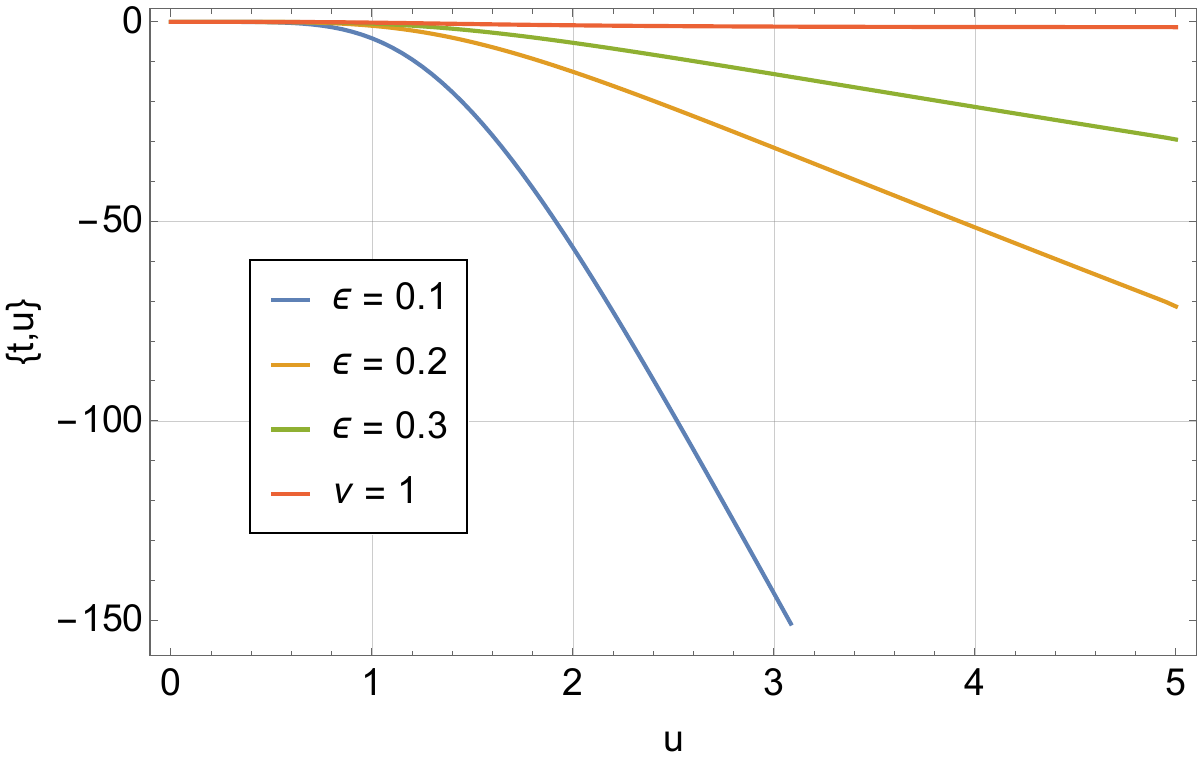}
        \label{fig_Sch_tanh}
      \end{minipage} 
    \end{tabular}
    \caption{Numerical results for \eqref{eq:tanh}, using the same values of $\epsilon$ as in Fig.\ \ref{fig_cubic_con}.}
    \label{fig_tanh}
  \end{figure}
  
Fig.\ \ref{fig_tanh} shows the results for
\begin{align}\label{eq:tanh}
g(u) = \tanh^3u
\end{align}
This corresponds to the case in which the boundary theory is smoothly coupled to the heat bath at $u=0$, and energy continues to be injected afterwards without ever turning off the coupling.
In this case as well, $t(u)$ approaches a constant, confirming the formation of a black hole. However, the late-time behavior of $\dot{t}$ does not match that of \eqref{BHsolution}, and $-\{t,u\}$, namely the black hole mass, continues to grow. Since dissipation is neglected in the high-temperature limit of the heat bath, it is natural that the energy keeps increasing.

\begin{figure}[t]
    \begin{tabular}{cc}
      \begin{minipage}[t]{0.45\hsize}
        \centering
        \subcaption{$t(u)$}
        \includegraphics[keepaspectratio, scale=0.35]{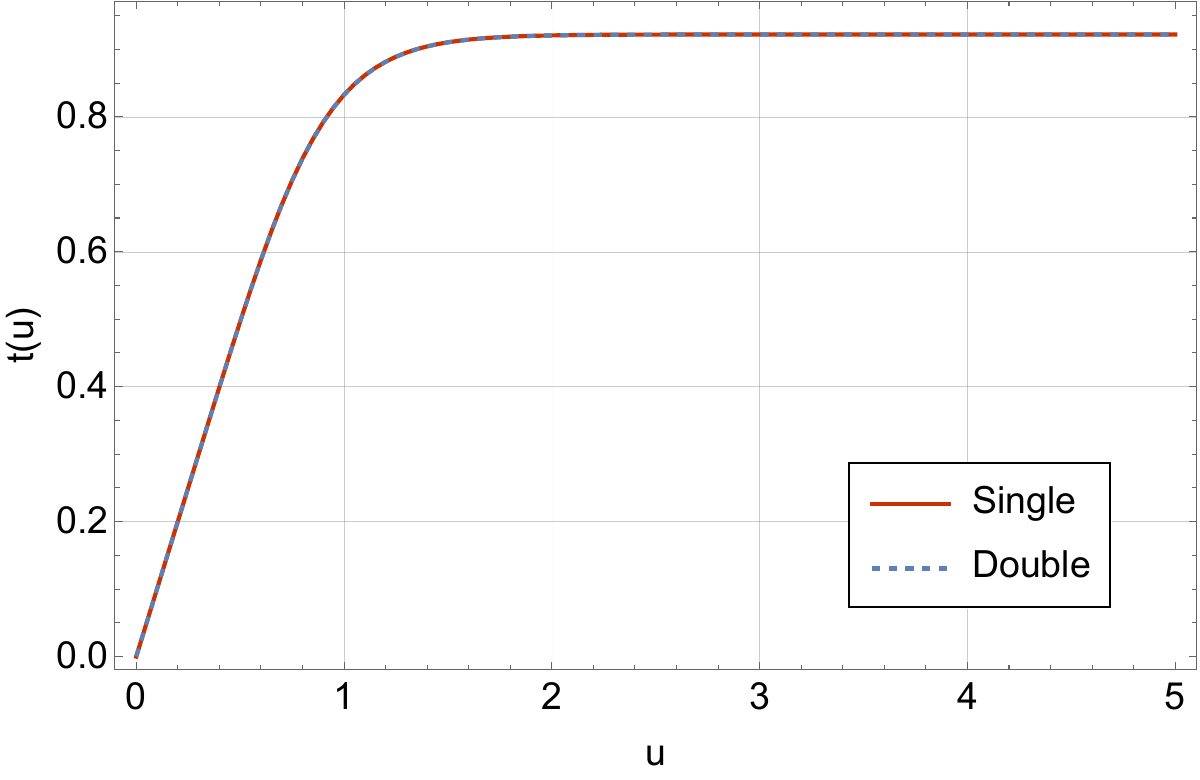}
        \label{fig_t_changer}
      \end{minipage} &
      \begin{minipage}[t]{0.45\hsize}
        \centering
        \subcaption{$\dot{t}(u)$}
        \includegraphics[keepaspectratio, scale=0.35]{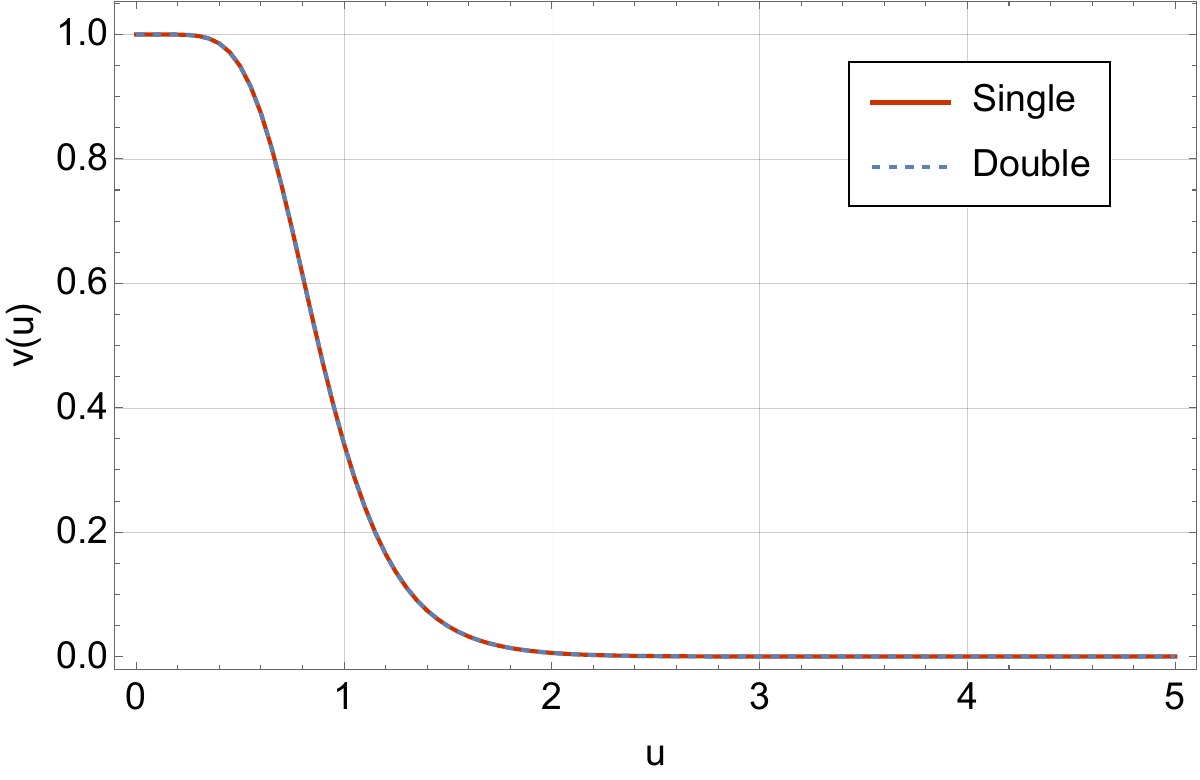}
        \label{fig_tdot_chenger}
      \end{minipage} \\[40pt]
   
      \begin{minipage}[t]{0.45\hsize}
        \centering
        \subcaption{$\log(\dot{t}(u))$}
        \includegraphics[keepaspectratio, scale=0.35]{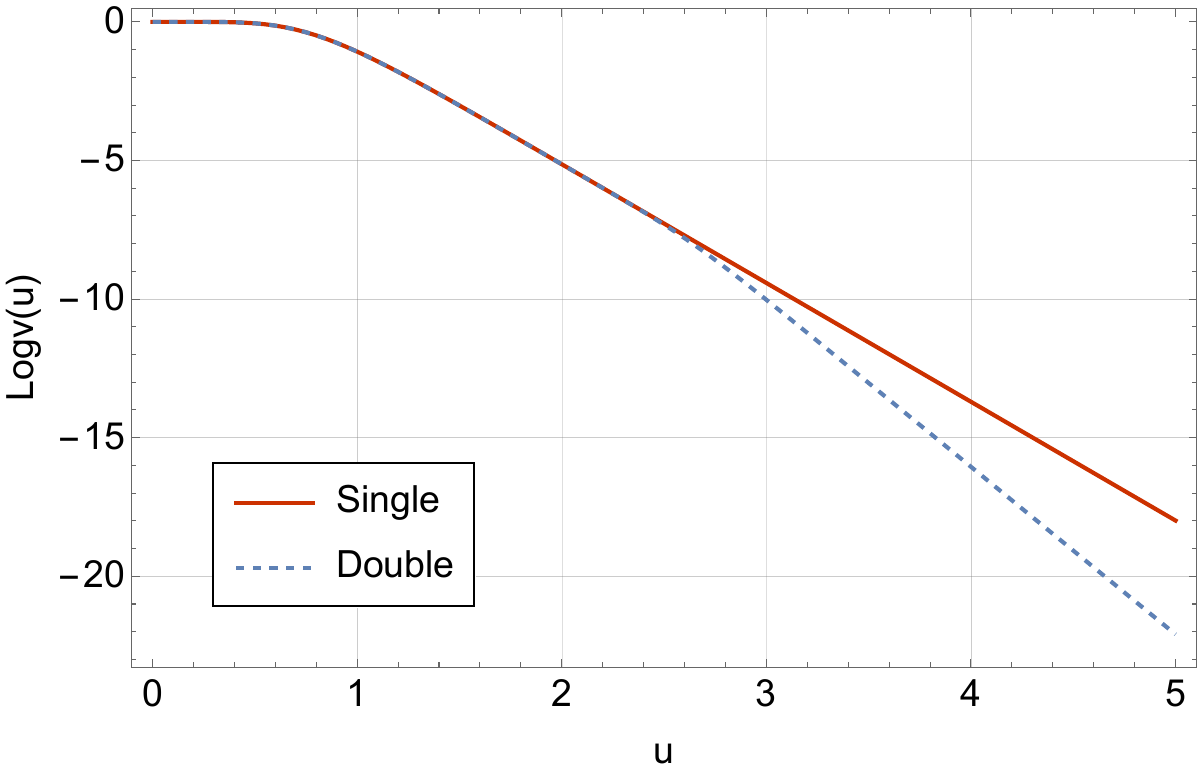}
        \label{fig_logtdot_changer}
      \end{minipage} &
      \begin{minipage}[t]{0.45\hsize}
        \centering
        \subcaption{$\{t,u\}$}
        \includegraphics[keepaspectratio, scale=0.35]{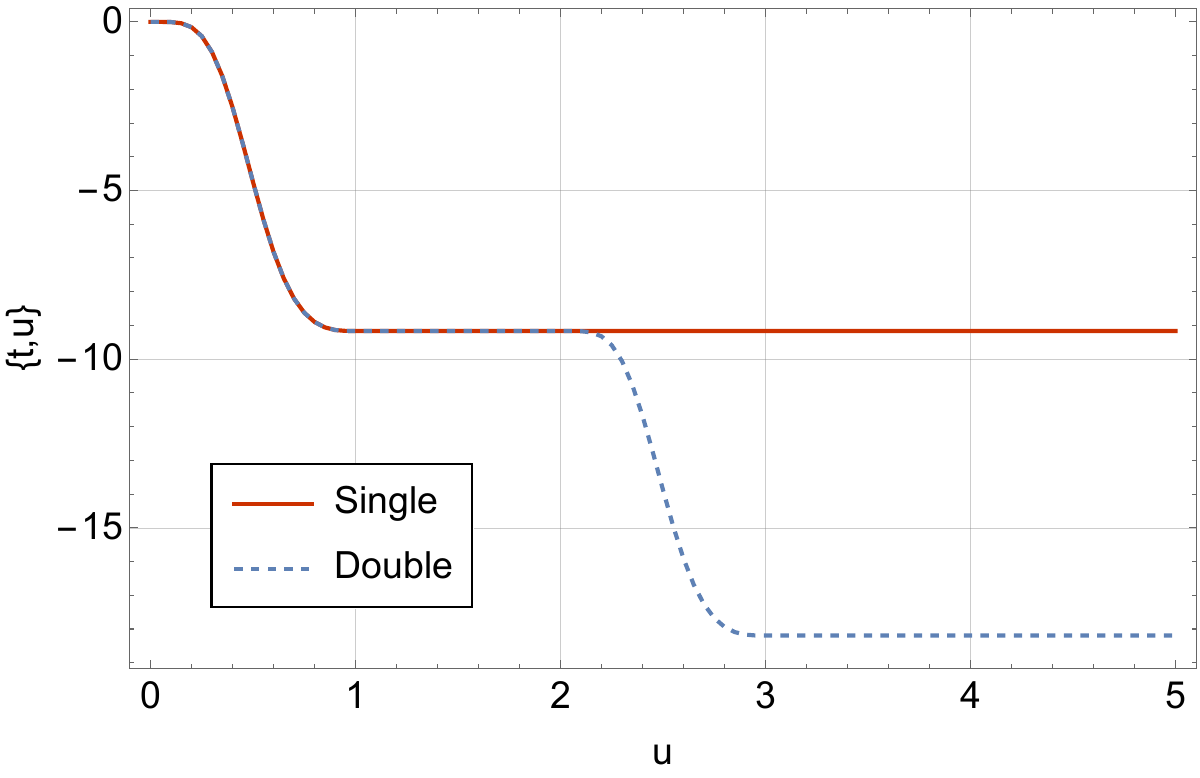}
        \label{fig_Sch_changer}
      \end{minipage} 
    \end{tabular}
    \caption{Single: Numerical results for \eqref{g_cubic}. Double: Numerical results for \eqref{eq:bath_changer}. Here, we set $\epsilon=0.2$.
    }
    \label{fig_bath_changer}
  \end{figure}

Fig.\ \ref{fig_bath_changer} shows the results obtained by choosing
\begin{align}
\label{eq:bath_changer}
g(u) = 
\begin{cases}
2^6u^3(1-u)^3&\text{ $u \in [0,1]$}\\
2^6(u-2)^3(3-u)^3&\text{ $u \in [2,3]$}\\
0\;&\text{(otherwise)}
\end{cases}
\end{align}
This corresponds to the case where the coupling between the boundary theory and the heat bath is turned off at $u=1$, turned on again at $u=2$, and finally turned off once more at $u=3$. During the second attachment $u\in[2,3]$, we inject additional heat into the black hole formed during the interval $u\in[0,1]$.
We see in Fig.\ \ref{fig_bath_changer} that the slope of $\log \dot t(u)$ and $-\{t,u\}$ (the black hole mass) increase again in the interval $u\in[2,3]$.

\section{Summary and discussion}\label{sec:Summery and discussion}
In this paper, we developed a prescription for describing the bulk dual of open quantum systems coupled to a heat bath within AdS/CFT, and applied it to JT gravity coupled to a scalar field. In the semiclassical and high-temperature limit, we numerically demonstrated that black-hole formation occurs.
The late-time behavior of the numerical solutions agrees with that of the black hole solution of JT gravity, in accordance with our expectation that the loss of purity in open-system dynamics provides a generic
mechanism for horizon formation in the bulk.
In the course of investigating black-hole formation, we also extended the holographic Lindblad prescription of \cite{Ishii:2025qpy} to the non-Markovian case. The detailed derivation of the extended dictionary, including dissipative effects, is given in Appendix \ref{app:bath AdS/CFT}.

In the calculations presented in this paper, we considered the high-temperature limit of the heat bath and neglected dissipation. As a result, as shown in Fig.\ \ref{fig_tanh}, the energy of the black hole continues to increase as long as the system remains coupled to the heat bath. If one retains the dissipative terms derived in Appendix \ref{app:bath AdS/CFT} without taking the high-temperature limit, it is natural to expect that fluctuations and dissipation will balance each other.
Then, without turning off the coupling to the heat bath, the system will thermalize into a stationary black hole having the same temperature as the bath. It may also be possible to induce black-hole evaporation by coupling the system to a cold heat bath as in \cite{Gaikwad:2022jar}. In that case as well, the dissipative terms cannot be neglected.

In the setup considered in this paper, black-hole formation was observed even when the boundary operator is coupled not to a dynamical bath but to an external source.
This observation is not surprising in itself, since horizon-bearing bulk geometries associated with unitary time evolution in isolated holographic systems have also been studied, as reviewed in Section \ref{sec:intro} (see also \cite{Louko:1998hc}).
In the context of JT gravity and the Sachdev--Ye--Kitaev (SYK) model, it has also been argued that certain pure states can be described by black-hole-like geometries \cite{Kourkoulou:2017zaj}.
Therefore, sufficiently microscopic observables may be needed to distinguish the black-hole formation found in the open-system case from that in the closed-system case.
Moreover, in the present work we have considered only saddle points with trivial topology.
In the quantum theory of JT gravity, it is known that summing over geometries with different topologies can drastically modify the late-time behavior compared with the result obtained from a fixed-topology saddle-point approximation \cite{Saad:2019lba,Saad:2019pqd}.
It therefore remains possible that differences between isolated and open systems may emerge from this perspective.

It is well known that black hole formation through gravitational collapse exhibits a universal critical phenomenon called Choptuik phenomenon \cite{Choptuik:1992jv}, which \cite{Dhar:2018pii} surveyed in the context of the low-energy Schwarzian description of a deformed SYK model and its JT gravity dual. 
It would be interesting to investigate whether a similar critical phenomenon emerges in our framework by varying the coupling profile $g(u)$.
Here, note that \cite{Dhar:2018pii} considers a unitary evolution with a Hamiltonian quenched by an external source, whereas our model is non-unitary.

Although we did not introduce sources in this paper because our focus was on the saddle point of the partition function, it is possible, as in \cite{Ishii:2025qpy}, to introduce sources and thereby express the generating functional of the boundary theory coupled to a heat bath in terms of the bulk theory as well. Since JT gravity has been studied extensively in connection with its duality to the SYK model and random matrix theory \cite{Maldacena:2016hyu,Maldacena:2016upp,Engelsoy:2016xyb,Jensen:2016pah,Saad:2019lba}, it would be interesting in future work to compare correlation functions in those theories coupled to a heat bath with corresponding calculations in open JT gravity.

\subsection*{Acknowledgement}
We thank Koji Hashimoto, Masazumi Honda, Takaaki Ishii, Satoshi Iso, Toshifumi Noumi, and Tomonori Ugajin for comments and discussions. R.A.\ would like to thank Seiken Chikazawa and Yasu Gen for helpful discussions. D.T.\ is supported by RIKEN Special Postdoctoral Researchers Program.

\appendix
\section{Coupling a heat bath to the boundary}\label{app:bath AdS/CFT}
Here, we derive the gravitational representation \eqref{genfnads} from the partition function \eqref{genfncft} of the CFT coupled to a heat bath.
First, by introducing auxiliary fields $\chi,\,\xi$ into \eqref{genfncft}, and defining $\mathcal{H}(x,y)\coloneq g(x)g(y)\eta(x-y)\Theta(x^0-y^0)$, the partition function can be rewritten as
\begin{align}
Z = &\int \mathcal{D}\varphi\mathcal{D}\chi\mathcal{D}\xi
\,e^{iI[C;\varphi]-i\int \mathrm{d}^d x \,\chi_r(\xi_r-O_r)-i\int \mathrm{d}^d x \,\chi_a (\xi_a-O_a)}\notag\\
&\times\exp\left(-\int \mathrm{d}^d x \mathrm{d}^d y\left[2i\xi_a(x)\mathcal{H}(x,y)\xi_r(y)
+ \frac{1}{2}\xi_a(x)\mathcal{N}(x,y)\xi_a(y)\right]\right)
\end{align}
Applying the GKPW relation \cite{Gubser:1998bc,Witten:1998qj} to this expression, we obtain
\begin{align}
    Z =& \int \mathbb{D}\Phi \mathcal{D}\chi\mathcal{D}\xi 
    \,e^{iS[M;\Phi,\chi]-i\int \mathrm{d}^d x(\chi_r\xi_r + \chi_a\xi_a)}\notag\\
    &\times \exp\left(-\int \mathrm{d}^d x \mathrm{d}^d y\left[2i\xi_a(x)\mathcal{H}(x,y)\xi_r(y) 
    + \frac{1}{2}\xi_a(x)\mathcal{N}(x,y)\xi_a(y)\right]\right).
    \label{genfnads with Aux field}
\end{align}
Here, the quantity $\chi$ appearing as an argument of the bulk action $S$ specifies the boundary condition for the scalar field $\Phi$:
\begin{equation}
    \Phi_{f/b}(z,x)\sim\left(\chi_a \pm \frac{\chi_r}{2}\right)z^{d-\Delta},\quad
    \Phi_E(z,x)\sim \mathcal{O}(z^{\Delta}), \quad(z\sim0).
\end{equation}
In what follows, we define the convolution product $\ast$ and the inner product $(\cdot,\cdot)$ of functions as follows
\begin{align}
    &\mathcal{G}\ast g (x)\coloneq \int \mathrm{d}^d y\, \mathcal{G}(x,y) g(y),\quad(g,h)\coloneq\int \mathrm{d}^d x\, g(x)h(x)
\end{align}
In \eqref{genfnads with Aux field}, after first performing the $\xi_r$ integral and then the $\chi_r$ integral, one obtains
\begin{align}
    &Z = \int  \mathcal{D}\chi_a\mathcal{D}\xi_a\Gamma[-2\mathcal{H}^T\ast\xi_a,\chi_a]
    e^{-i(\chi_a,\xi_a)-(\xi_a,\mathcal{N}\ast\xi_a)/2},\\
    &\Gamma[\chi_r,\chi_a]\coloneq\int\mathbb{D}\Phi\exp(iS[M;\Phi,\chi])
\end{align}
where $\mathcal{H}^T(x,y)\coloneq \mathcal{H}(y,x)$.
Completing the square in the exponent and shifting the integration variable from $\xi_a$ to $\xi_a - i\hat{\mathcal{N}} \ast \chi_a$, one obtains
\begin{equation}
    Z = \int \mathcal{D}\chi_a\mathcal{D}\xi_a\,\Gamma[2i\mathcal{H}^T\ast(\hat{\mathcal{N}}\ast\chi_a)-2\mathcal{H}^T\ast\xi_a,\chi_a]
    e^{-(\chi_a,\hat{\mathcal{N}}\ast\chi_a)/2-(\xi_a,\mathcal{N}\ast\xi_a)/2}
    \label{Z generic}
\end{equation}
where we used $\mathcal{N}(x,y) = \mathcal{N}(y,x)$. Equation \eqref{Z generic} is a generalization of \eqref{genfnads} that includes the case in which dissipation cannot be neglected. In particular, when $\mathcal{H}$ can be neglected in the high-temperature limit, it reduces to \eqref{genfnads}.

\subsection{How to design a bath to reproduce \texorpdfstring{\eqref{normalized_delta}}{(normalizeddelta)}}
For JT gravity, the boundary theory is one-dimensional, and thus we can use Feynman--Vernon or Caldeira--Leggett model \cite{FEYNMAN1963118, CaldeiraLeggett} for bath, which consists of an infinite number of harmonic oscillators.
In that model, $\eta$ and $\nu$ are given as
\begin{align}\label{eq: Caldeira Leggett}
	\eta(s) = -\int_0^\infty \frac{\d \omega}{2\pi}J(\omega)\sin\left(\omega s \right),\qquad
	\nu(s) = \int_0^\infty \frac{\d \omega}{2\pi}J(\omega)\coth\left(\frac{\omega \beta}{2} \right) \cos \left(\omega s \right),
\end{align}
where $J$ is a spectral function that we can choose.
It is obvious that $\nu$ is dominant in the high-temperature limit.

To reproduce \eqref{normalized_delta}, we design the bath spectrum as
\begin{align}
    J(\omega) = \gamma \omega e^{-\epsilon \omega}\qquad
    (\mbox{$\gamma, \epsilon$: positive constants}).
\end{align}
Then, we find
\begin{align}
    \nu(u) \simeq \frac{\gamma}{\beta}\cdot \frac{\epsilon}{\pi}\frac{1}{u^2 + \epsilon^2}
    \qquad
    (\beta\to +0),
\end{align}
which is nothing but \eqref{normalized_delta}, up to a normalization constant.

\section{Integrating out scalar field}\label{app:On-shell action}
Here, we explain the derivation of \eqref{on-shellaction}. If $\det(\nabla^2-m^2)$ is neglected, integrating out the scalar field is equivalent to evaluating the on-shell action for the matter field. A similar computation in JT gravity is already performed in \cite{Arias:2023ygy}.

We first summarize the matching conditions \cite{Skenderis:2008dh,Skenderis:2008dg} that the metric and the scalar field must satisfy on each gluing surface in Fig.\ \ref{fig_bulk_contour}. Roughly speaking, the condition on the metric is that the manifold should be glued smoothly, and in the present case this can be achieved by choosing the gluing surfaces to be geodesics given by constant-$t$ slices. Moreover, by using a diffeomorphism that leaves the Poincar\'e metric invariant, one may choose the gluing surface between Euclidean and Lorentzian segments (green and red in Fig.\ \ref{fig_bulk_contour}) to be given by,
\begin{align}\label{eq:gauge1}
    \tau_f = t_f = t_b = \tau_b = 0
\end{align}
and the surface between the two Lorentzian segments (yellow in Fig.\ \ref{fig_bulk_contour}) by
\begin{align}\label{eq:gauge2}
    t_f = t_b = t_c.
\end{align}
Here, $f$ ($b$) indicates forward (backward) as in the main text, and $t$ ($\tau$) represents Lorentzian (Euclidean) Poincar\'e time.
In addition, $t_c$ is an arbitrary positive number.
The matching conditions for the scalar field can then be written as follows: for any $z$,
\begin{align}
    &\Phi_{E,f}(\tau_f = 0) = \Phi_f(t_f=0),\qquad i\partial_\tau \Phi_{E,f}(\tau_f = 0) = \partial_t \Phi_f(t_f = 0),\\
    &\Phi_{E,b}(\tau_b = 0) = \Phi_b(t_b=0),\qquad i\partial_\tau \Phi_{E,b}(\tau_b = 0) = \partial_t \Phi_b(t_b = 0),\\
    &\Phi_{f}(t_f = t_c) = \Phi_b(t_b = t_c),\qquad \partial_t \Phi_{f}(t_f = t_c) = \partial_t \Phi_b(t_b = t_c).
\end{align}
Here, $\Phi_{E,f}$ and $\Phi_{E,b}$ denote the scalar fields on the forward and backward Euclidean segments, respectively.

To derive the on-shell action, one must solve the equation of motion $(\nabla^2-m^2)\Phi = 0$ subject to both the above matching conditions and the boundary condition \eqref{bdy_cond_scaler}. Such a solution can be expressed in terms of the bulk-to-boundary propagator in two dimensions as follows:
\begin{align}
    &\Phi_{Ef,b}(z,\tau) = i\int_0^{t_c} \mathrm{d}\tilde t \,K_\Delta(z,\tau-i\tilde t)(\tilde \lambda_f(\tilde t) - \tilde \lambda_b(\tilde t))\\
    &\Phi_f(z,t) = \int_0^{t_c}\mathrm{d}\tilde t\, K_\Delta^-(z,t-\tilde t)\tilde\lambda_f(\tilde t)
                   -\int_0^{t_c}\mathrm{d}\tilde t \, N_\Delta^+(z,t-\tilde t)\tilde\lambda_b(\tilde t)\label{scaler classical solution f}\\
    &\Phi_b(z,t) = \int_0^{t_c}\mathrm{d}\tilde t \, K_\Delta^+(z,t-\tilde t)\tilde\lambda_b(\tilde t)
                   -\int_0^{t_c}\mathrm{d}\tilde t \, N_\Delta^-(z,t-\tilde t)\tilde\lambda_f(\tilde t)\label{scaler classical solution b}\\
    &K_\Delta(z,\tau)\coloneq C_\Delta\left(\frac{z}{z^2 + \tau^2}\right)^\Delta,\qquad
    K_\Delta^{\pm} (z,t)
    \coloneq \mp iC_\Delta\left(\frac{z}{z^2-(e^{\pm i\varepsilon}t)^2}\right)^\Delta,\\
    &N_\Delta^{\pm}(z,t)
    \coloneq\pm iC_\Delta \left(\frac{z}{z^2-(t\pm i\varepsilon)^2}\right)^\Delta.
\end{align}
Here, $K_\Delta^\pm$ and $N_\Delta^\pm$ are analytic continuations of $K_\Delta$, and $\varepsilon$ is an infinitesimal constant specifying from which side of the branch cut the propagator is evaluated.\footnote{
We choose the branch of the power function $w^\Delta$ such that $-\pi < \arg w<\pi$.
}
Inside the integrals, they behave as
\begin{align}
K_\Delta^\pm(z,t)\sim z^{1-\Delta}\delta(t),\qquad N_\Delta^\pm(z,t)\sim \mathcal{O}(z^\Delta)\qquad(z\sim0)
\end{align}
so that the former contains the non-normalizable mode, while the latter is purely normalizable.
Moreover, $\tilde\lambda_{f,b}(t_{f,b})$ denotes the source defined with respect to the Poincar\'e time coordinate $t$, and it must be distinguished from the source $\lambda(u)$ defined in terms of the boundary time $u$. Since the boundary condition \eqref{bdy_cond_JT} implies $z(u) = \epsilon \dot{t}(u) + \mathcal{O}(\epsilon^2)$, the relation between them is given by
\begin{align}
z^{1-\Delta}\tilde \lambda_{f,b}(t_{f,b}(u)) = \epsilon^{1-\Delta}\lambda(u)\implies\tilde \lambda_{f,b}(t_{f,b}(u)) =\dot t_{f,b}^{\Delta-1}(u)\lambda(u).
\end{align}

Substituting these expressions into \eqref{action_scaler} together with the counterterm, one obtains the on-shell action. In what follows, we assume $1/2<\Delta<3/2$. Since no source is turned on in the Euclidean segments, the on-shell action vanishes there. For the Lorentzian segments, the bulk contribution to the action vanishes on shell, and one finds
\begin{align}
    S_{m,\mathrm{eff}}\coloneq(S_{m} + S_{ct})|_{\mathrm{on-shell}} \sim\frac{1}{2}\int_{\partial M}\mathrm{d}t\left(\Phi\partial_z\Phi +(\Delta-1)\frac{1}{z}\Phi^2\right)\qquad(\epsilon \sim 0).\label{scaler on-shell formal}
\end{align}
We have not written the boundary terms on the gluing surfaces explicitly, since they vanish by virtue of the matching conditions.

To evaluate this in the limit $\epsilon \to 0$, we expand \eqref{scaler classical solution f} and \eqref{scaler classical solution b} near $z = 0$. Under the assumption $1/2<\Delta<3/2$, the relevant terms are
\begin{align}
    \Phi_f &= iC_{\Delta}z^{\Delta}\left(
    -\mathrm{Pf}\int_0^{t_c} \mathrm{d}\tilde t\, 
    \frac{\tilde \lambda_b(\tilde t)
    e^{i\pi\Delta \mathrm{sgn}(t-\tilde{t})}}{|t-\tilde t|^{2\Delta }} 
    +\mathrm{Pf}\int_0^{t_c} \mathrm{d}\tilde t\,
    \frac{\tilde \lambda_f(\tilde t)
    e^{-i\pi\Delta}}{|t-\tilde t|^{2\Delta }}  \right)\notag\\
    &~ +  z^{1-\Delta } 
    \tilde \lambda_f(t)+ o(z^{\Delta})
     \label{series expansion of scaler}
\end{align}
Here, Pf denotes the Hadamard regularization of the integral. For an integral with a singularity at $x=a$, if the function $f$ admits a Taylor expansion around $a$, it is defined by
\begin{equation}
    \mathrm{Pf}\int_a^b dx \frac{f(x)}{(x-a)^\alpha} 
    \coloneq \lim_{\delta\to+0}\left[\int_{a+\delta}^b\frac{f(x)}{(x-a)^\alpha} + \sum_{n+1-\alpha<0}\frac{f^{(n)}(a)}{n!}\frac{\delta^{n+1-\alpha}}{n+1-\alpha}\right].
\end{equation}
When the singular point is in the interior of the integration range, we devide it into two intervals at the point and apply the above regularization separately. The expansion of $\Phi_b$ is obtained by interchanging $f$ and $b$ in the expansion of $\Phi_f$ and then taking the complex conjugate.

If we formally write the asymptotic expansion of $\Phi$ as
\begin{align}
    \Phi =  C_0(t)z^{\Delta}+D_0(t)z^{1-\Delta} + o(z^{\Delta})
\end{align}
the on-shell action \eqref{scaler on-shell formal}, under the assumption $1/2<\Delta<3/2$, reduces to
\begin{align}
    S_{m,\mathrm{eff}} = \left(\Delta-\frac{1}{2}\right)\int_{\partial M} \mathrm{d}t\, C_0(t)D_0(t).
\end{align}
Substituting \eqref{series expansion of scaler} and the corresponding expression for $\Phi_b$ into this expression then yields \eqref{on-shellaction}.

\section{Derivation of semiclassical equation of motion}\label{app:eom}
Here, we explain the derivation of \eqref{eom}, focusing in particular on the variation of $W$. In the discussion below, we omit the overall factor $-(\Delta-1/2)C_\Delta$ in $iS_{m,\mathrm{eff}}$. For simplicity, we also assume $1/2<\Delta<1$, which is satisfied by our choice in section \ref{sec:Numerical verification}.

First, from the expression \eqref{on-shellaction}, one finds that $S_{m,\mathrm{eff}}[t,t] =0$ when $t_f = t_b = t$. Using this fact, one can vary $W$ with respect to $t_f$ and then compute the average over $\lambda$, obtaining
\begin{align}
    &\delta_{t_f}(iW[t_f,t_b])|_{t_f=t_b =t} \notag\\
    &= \frac{\int\mathcal{D}\lambda 
    \exp\left(-\frac{1}{2}\int \mathrm{d}u_1 \int \mathrm{d}u_2 \,\lambda(u_1)\hat{\mathcal{N}}(u_1,u_2)\lambda(u_2) + iS_{m,\mathrm{eff}}[t,t]\right)\delta_{t_f} (iS_{m,\mathrm{eff}}[t_f,t])|_{t_f = t}}
    {\int \mathcal{D}\lambda \exp\left(-\frac{1}{2}\int \mathrm{d}u_1 \int \mathrm{d}u_2 \,\lambda(u_1)\hat{\mathcal{N}}(u_1,u_2)\lambda(u_2)\right)}\notag\\
    & = \delta_{t_f}\left(\int_0^\infty \mathrm{d}u_1 \mathrm{Pf}\int_0^\infty \mathrm{d}u_2 \,\mathcal{N}(u_1,u_2)\left[
    e^{-i\pi \Delta}\frac{(\dot t_f(u_1)\dot t_f(u_2))^\Delta}{|t_f(u_1)-t_f(u_2)|^{2\Delta}}
    \right.\right.\notag\\
    & \hspace{60pt}\left.\left.\left.
    +e^{i\pi \Delta}\frac{(\dot t(u_1)\dot t(u_2))^\Delta}{|t(u_1)-t(u_2)|^{2\Delta}}
    -2e^{i\pi \Delta\mathrm{sgn}(t_f(u_1)-t(u_2))}
    \frac{(\dot t_f(u_1)\dot t(u_2))^\Delta}{|t_f(u_1)-t(u_2)|^{2\Delta}}
    \right]\right)\right|_{t_f = t}.
\end{align}
Here, we have used $iS_{m,\mathrm{eff}}[t,t] = 0$.
Thus, under our assumption, computing the variation of $W$ is equivalent to computing the variation of the effective action
\begin{align}
    &iW_*[t_f,t_b]\notag \\
    &=\int_0^\infty \mathrm{d}u_1 \mathrm{Pf}\int_0^\infty \mathrm{d}u_2 \,\mathcal{N}(u_1,u_2)\left[
    e^{-i\pi \Delta}\frac{(\dot t_f(u_1)\dot t_f(u_2))^\Delta}{|t_f(u_1)-t_f(u_2)|^{2\Delta}}
    \right.\notag\\
    & \qquad\qquad\left.
    +e^{i\pi \Delta}\frac{(\dot t_b(u_1)\dot t_b(u_2))^\Delta}{|t_b(u_1)-t_b(u_2)|^{2\Delta}}
    -2e^{i\pi \Delta\mathrm{sgn}(t_f(u_1)-t_bt(u_2))}
    \frac{(\dot t_f(u_1)\dot t_b(u_2))^\Delta}{|t_f(u_1)-t_b(u_2)|^{2\Delta}}
    \right].\label{W effective}
\end{align}
The second term in \eqref{W effective} does not contain $t_f$, and therefore does not contribute to the following calculation. For convenience, we define
\begin{align}
    &K_{ff}(u_1,u_2)\coloneq
    \mathcal{N}(u_1,u_2)\frac{(\dot t_f(u_1)\dot t_f(u_2))^\Delta}{|t_f(u_1)-t_f(u_2)|^{2\Delta}},
    \\
    &K_{fb}(u_1,u_2) \coloneq 
    \mathcal{N}(u_1,u_2)\frac{(\dot t_f(u_1)\dot t_b(u_2))^\Delta}{|t_f(u_1)-t_b(u_2)|^{2\Delta}}.
\end{align}

Note that the Hadamard regularization, $\mathrm{Pf}$, appears in $W_*$. To carefully carry out the calculation, it is convenient to rewrite $W_*$ according to the definition of $\mathrm{Pf}$.
The first and third term in \eqref{W effective} are respectively written as
\begin{align}
    &I_{ff} [t_f]=\int_0^\infty \mathrm{d}u_1 \lim_{\varepsilon\to +0}e^{-i\pi\Delta}\left[\left(\int_0^{u_1-\varepsilon} + \int_{u_1 + \varepsilon}^\infty\right)\mathrm{d}u_2\,K_{ff}(u_1,u_2) +\mathcal{N}(u_1,u_1)\frac{2\varepsilon^{1-2\Delta}}{1-2\Delta}\right] ,\label{Iff}\\
    &I_{fb} [t_f,t_b]=-2\int_0^\infty \mathrm{d}u_1 \lim_{\varepsilon\to +0}\left[\left(e^{i\pi\Delta}\int_0^{u_*-\varepsilon} + e^{-i\pi\Delta}\int_{u_* + \varepsilon}^\infty\right)\mathrm{d}u_2\,K_{fb}(u_1,u_2) \right.\notag\\
    &\hspace{160pt}\left.
    +2\cos(\pi\Delta)\mathcal{N}(u_1,u_*)\left(\frac{\dot t_f (u_1)}{\dot t_b (u_*)}\right)^\Delta\frac{\varepsilon^{1-2\Delta}}{1-2\Delta}\right].
\end{align}
Here, $u_*$ is the time defined by $t_f(u_1) = t_b(u_*)$. Since $t_b$ is the function mapping boundary time to the Poincaré time coordinate and is monotonically increasing, such a $u_*$ is uniquely determined once $t_f$, $t_b$, and $u_1$ are specified.

In the following calculation, we impose $\delta t(0) = \delta t(\infty) = 0$. 
This condition comes from the gauge conditions \eqref{eq:gauge1} and \eqref{eq:gauge2}.

\paragraph{Variation of $I_{ff}$}
We begin with the variation of $I_{ff}$.
\begin{align}
    \delta_{t_f}K_{ff} (u_1,u_2)= \Delta K_{ff}(u_1,u_2)\left(\frac{\delta \dot t_f(u_1)}{\dot t_f(u_1)}+\frac{\delta \dot t_f(u_2)}{\dot t_f(u_2)}-2\frac{\delta t_f(u_1)-\delta t_f(u_2)}{t_f(u_1)-t_f(u_2)}\right).\label{delta Kff}
\end{align}
For \eqref{Iff}, the counterterm does not depend on $t_f$ and therefore drops out under the variation. Accordingly, the variation is therefore given by integrating $\delta_{t_f}K_{ff}$. Taking into account that the integration range of the $u_2$ depends on $u_1$, one obtains
\begin{align}
    &\delta_{t_f}I_{ff}|_{t_f = t} 
    = e^{-i\pi\Delta}\int_0^\infty \mathrm{d}u_1 
    \lim_{\varepsilon \to +0}
    \left[-2\left(\int_0^{u_1-\varepsilon} + \int_{u_1 + \varepsilon} ^\infty\right) \mathrm{d}u_2 \, 
    \delta t_f(u_1)\mathcal{A}(u_1,u_2)\right.\notag\\
    &\qquad
    + \Delta\left( \delta t_f(u_1-\epsilon)\frac{K_{t}(u_1,u_1-\varepsilon)}{\dot t(u_1-\epsilon)}
    -  \delta t_f(u_1+\epsilon)\frac{K_{t}(u_1,u_1+\varepsilon)}{\dot t(u_1+\epsilon)}\right.\notag\\
    &\hspace{120pt}\left.\left.
    -\delta t_f(u_1)\frac{K_t(u_1,u_1-\varepsilon)}{\dot t(u_1)}
    +\delta t_f(u_1)\frac{K_t(u_1,u_1+\varepsilon)}{\dot t(u_1)}
    \right)\right],\label{variation of Iff: unsimplified}\\
    &\mathcal{A}(u_1,u_2)\coloneq 
    2\Delta\frac{K_t(u_1,u_2)}{t(u_1)-t(u_2)}
    + \Delta\frac{\partial}{\partial u_1}\left(\frac{K_t(u_1,u_2)}{\dot t(u_1)} \right)
    \label{def of A}
\end{align}
Here, $K_t$ denotes the expression obtained from $K_{ff}$ by setting $t_f\to t$. We have also used the following relation for a general function $\mathcal{G}(u_1,u_2)$:
\begin{align}
    \int_0^\infty \mathrm{d}u_1 \int_0^{u_1-\varepsilon} \mathrm{d}u_2\, \delta t_f(u_2)\mathcal{G}(u_1,u_2)
     = \int_0^\infty \mathrm{d}u_1 \,\delta t_f(u_1) \int^\infty_{u_1+\varepsilon}\mathrm{d}u_2 \,\mathcal{G}(u_2,u_1).
\end{align}
The last terms in \eqref{variation of Iff: unsimplified} arise from the partial integrations containing $\delta \dot t_f$. Combining these boundary terms to write $B_\varepsilon(u_1)$, we find
\begin{equation}
    B_\varepsilon(u_1) = 2e^{-i\pi\Delta}\Delta\mathcal{N}(u_1,u_1)\varepsilon^{1-2\Delta}\frac{\mathrm{d}}{\mathrm{d}u_1}\frac{\delta t_f(u_1)}{\dot t_f(u_1)} + O(\varepsilon^{2-2\Delta}).
\end{equation}
Therefore, the variation of $I_{ff}$ can be written as
\begin{align}
    &\delta_{t_f}I_{ff}|_{t_f = t} = -2e^{-i\pi\Delta}\int_0^\infty \mathrm{d}u_1 \,\delta t_f(u_1)\notag\\
    &\times\lim_{\varepsilon \to +0}\left[\left(\int_0^{u_1-\varepsilon} + \int_{u_1 + \varepsilon} ^\infty \right) \mathrm{d}u_2 \,\mathcal{A}(u_1,u_2)
    -\Delta\varepsilon^{1-2\Delta}
    \frac{1}{\dot t_f(u_1)}
    \frac{\mathrm{d}}{\mathrm{d}u_1}
    \mathcal{N}(u_1,u_1)
    \right].\label{variation of Iff}
\end{align}

\paragraph{Variation of $I_{fb}$}
Next, we consider the variation of $I_{fb}$. First, the variation of $K_{fb}$ is given by
\begin{equation}
    \delta_{t_f} K _{fb}(u_1,u_2) = \Delta K_{fb}(u_1,u_2)
    \left(\frac{\delta\dot t_f(u_1)}{\dot t_f(u_1)} - 2\frac{\delta t_f(u_1)}{t_f(u_1)-t_b(u_2)}\right).
\end{equation}
Unlike the case of $I_{ff}$, the counterterm of $I_{fb}$ depends on $t_f$. In particular, $u_*$ itself depends on $t_f$. From $t_f(u_1) = t_b(u_*)$, its variation with respect to $t_f$ is given by
\begin{align}
    \dot t_b(u_*)\delta_{t_f} u_* = \delta_{t_f} t_f(u_1) \implies \delta_{t_f} u_*|_{t_f = t_b = t} =\frac{\delta t_f(u_1)}{\dot t(u_1)} .
\end{align}
Using this, the variation of the counterterm can be computed as
\begin{align}
    &\left.\delta_{t_f}\left(\mathcal{N}(u_1,u_*)\left(\frac{\dot t_f (u_1)}{\dot t_b(u_*)}\right)^\Delta\right)\right|_{t_f = t_b = t} \notag\\
    &=
   \Delta \frac{\mathrm{d}}{\mathrm{d}u_1} \left(\mathcal{N}(u_1,u_1)\frac{\delta t_f (u_1)}{\dot t(u_1)}\right) 
    -\left(\Delta- \frac{1}{2}\right) \frac{\delta t_f (u_1)}{\dot t(u_1)}\frac{\mathrm{d}}{\mathrm{d}u_1}\mathcal{N}(u_1,u_1)
\end{align}
where we have used $\partial_{u_2}\mathcal{N}|_{u_1 = u_2 = u} = 1/2 \cdot \mathrm{d}\mathcal{N}(u,u)/\mathrm{d}u$, which follows from $\mathcal{N}(u_1,u_2) = \mathcal{N}(u_2,u_1)$. Using these expressions, the variation can be written as
\begin{align}
    &\delta_{t_f}I_{fb}|_{t_f = t_b = t} = 
    -2\int_0^\infty \mathrm{d}u_1 \lim_{\varepsilon\to +0}
    \left[
    \left(
    e^{i\pi\Delta}\int_0^{u_1-\varepsilon} + e^{-i\pi\Delta}\int_{u_1 + \varepsilon}^\infty
    \right)
    \mathrm{d}u_2\,\delta_{t_f} K_{fb}(u_1,u_2)|_{t_f = t_b = t} 
    \right.\notag\\
    &\hspace{120pt}\left. 
    +(e^{i\pi\Delta}K_{t}(u_1,u_1-\epsilon)-e^{-i\pi\Delta}K_t(u_1,u_1+\epsilon))\delta_{t_f} u_*
    \right.\notag\\
    &\hspace{120pt}\left.
    +2\cos(\pi\Delta)\frac{\varepsilon^{1-2\Delta}}{1-2\Delta}\left.\delta_{t_f}\left(\mathcal{N}(u_1,u_*)\left(\frac{\dot t_f (u_1)}{\dot t_b(u_*)}\right)^\Delta\right)\right|_{t_f = t_b = t}
    \right].
\end{align}
One can rearrange this as
\begin{align}
    &\delta_{t_f}I_{fb}|_{t_f = t_b = t} = 
    2\int_0^\infty \mathrm{d}u_1 \lim_{\varepsilon\to +0}
    \left[
    \left(
    e^{i\pi\Delta}\int_0^{u_1-\varepsilon} + e^{-i\pi\Delta}\int_{u_1 + \varepsilon}^\infty
    \right)
    \mathrm{d}u_2 \, \delta t_f(u_1)\mathcal{A}(u_1,u_2)\right.\notag\\
    &\hspace{90pt}\left.
     + (\Delta-1)\frac{\delta t_f(u_1)}{\dot t (u_1)}\left(e^{i\pi\Delta}K_t(u_1,u_1-\varepsilon) - e^{-i\pi\Delta}K_t(u_1,u_1+\varepsilon)\right)\right.\notag\\
    &\hspace{90pt}\left.
    -\cos(\pi\Delta)\varepsilon^{1-2\Delta} \frac{\delta t_f (u_1)}{\dot t(u_1)}
    \frac{\mathrm{d}}{\mathrm{d}u_1}\mathcal{N}(u_1,u_1)
    \right].
\end{align}
For the boundary term, one finds the expansion
\begin{align}
    &e^{i\pi\Delta}K_t(u_1,u_1-\varepsilon) - e^{-i\pi\Delta}K_t(u_1,u_1+\varepsilon)\notag\\
    &=2i\sin(\pi\Delta)\mathcal{N}(u_1,u_1)\varepsilon^{-2\Delta}-\cos(\pi\Delta)\frac{\mathrm{d}}{\mathrm{d}u_1}\mathcal{N}(u_1,u_1)\varepsilon^{1-2\Delta} + O(\varepsilon^{2-2\Delta}).
\end{align}
Therefore, retaining only the terms that survive in the limit $\varepsilon \to 0$, one finds
\begin{align}
    &\delta_{t_f}I_{fb}|_{t_f = t_b = t} = 
    2\int_0^\infty \mathrm{d}u_1 \,\delta t_f(u_1) \lim_{\varepsilon\to +0}
    \left[
    \left(
    e^{i\pi\Delta}\int_0^{u_1-\varepsilon} + e^{-i\pi\Delta}\int_{u_1 + \varepsilon}^\infty
    \right)
    \mathrm{d}u_2 \,\mathcal{A}(u_1,u_2)\right.\notag\\
    &\hspace{10pt}\left.
    -2i(1-\Delta)\sin(\pi\Delta)\varepsilon^{-2\Delta}\frac{\mathcal{N}(u_1,u_1)}{\dot t(u_1)}
     - \Delta\cos(\pi\Delta)\varepsilon^{1-2\Delta}\frac{1}{\dot t(u_1)}\frac{\mathrm{d}}{\mathrm{d}u_1}\mathcal{N}(u_1,u_1)
    \right].
\end{align}

\paragraph{Final result}
Collecting the above results, the variation of $W_*$ takes the form:
\begin{align}
    &\delta_{t_f}(iW_*)|_{t_f = t_b = t} = 4i\sin(\pi\Delta)\int_0^\infty \mathrm{d}u_1\,\delta t_f(u_1)\lim_{\varepsilon \to +0} \left(\int_0^{u_1-\varepsilon}\mathrm{d}u_2\,\mathcal{A}(u_1,u_2)  + C(\varepsilon;u_1)\right),\label{variation of W*}\\
    &C(\varepsilon;u_1) \coloneq -(1-\Delta)\varepsilon^{-2\Delta}\frac{\mathcal{N}(u_1,u_1)}{\dot t(u_1)}
     -\frac{\Delta}{2}\varepsilon^{1-2\Delta}\frac{1}{\dot t(u_1)}\frac{\mathrm{d}}{\mathrm{d}u_1}\mathcal{N}(u_1,u_1).\label{ct term in eom}
\end{align}
In particular, setting $u_2 = u_1 + x$ and expanding $\mathcal{A}(u_1,u_2)$ around $x=0$, one finds
\begin{align}
    \mathcal{A}(u_1,u_2) =& -2\Delta(1-\Delta)\frac{\mathrm{sgn}(x)}{|x|^{2\Delta + 1}}
    \frac{\mathcal{N}(u_1,u_1)}{\dot t (u_1)} \notag\\
    &+ \Delta\left(\Delta-\frac{1}{2}\right)\frac{1}{|x|^{2\Delta}\dot t(u_1)}\frac{\mathrm{d}}{\mathrm{d}u_1}\mathcal{N}(u_1,u_1) + O(|x|^{1-2\Delta})
\end{align}
so that the integral appearing in \eqref{variation of W*} can be expanded as
\begin{align}
    \int_0^{u_1-\varepsilon}\mathrm{d}u_2\,\mathcal{A}(u_1,u_2) = (1-\Delta)\frac{\mathcal{N}(u_1,u_1)}{\dot t (u_1)} \varepsilon^{-2\Delta} + \frac{\Delta}{2} \frac{1}{\dot t (u_1)}\frac{\mathrm{d}}{\mathrm{d}u_1}\mathcal{N}(u_1,u_1) + O(\varepsilon^0).
\end{align}
Thus, \eqref{ct term in eom} provides the correct counterterm needed to cancel the divergence. Therefore, the limit in \eqref{variation of W*} may be rewritten in terms of the Hadamard regularization. Combining this result with
\begin{align}
    \delta_{t_f} S_\mathrm{Sch}[t_f]|_{t_f = t} =\frac{1}{8\pi G} \int \mathrm{d}u\frac{\delta t_f(u)}{\dot t(u)}\frac{\mathrm{d}}{\mathrm{d}u}\{t,u\}
\end{align}
one obtains the equation of motion \eqref{eom}.

So far, we have performed the variation carefully, keeping the definition of \(\mathrm{Pf}\) explicit.  However, the same result can be obtained by a formal variation in which one temporarily ignores the finite-part prescription and restores it at the end.  This is because \(\mathrm{Pf}\) defines the analytic continuation in \(\Delta\): one may first perform the computation in the convergence domain \(2\Delta+1<1\), where the integral is well defined without the finite-part prescription, and then analytically continue the result to the range \(1/2<\Delta<1\) considered in this paper.

\section{Numerical method and convergence}
\begin{figure}[t]
    \begin{tabular}{cc}
      \begin{minipage}[t]{0.45\hsize}
        \centering
        \subcaption{$t(u)$}
        \includegraphics[keepaspectratio, scale=0.35]{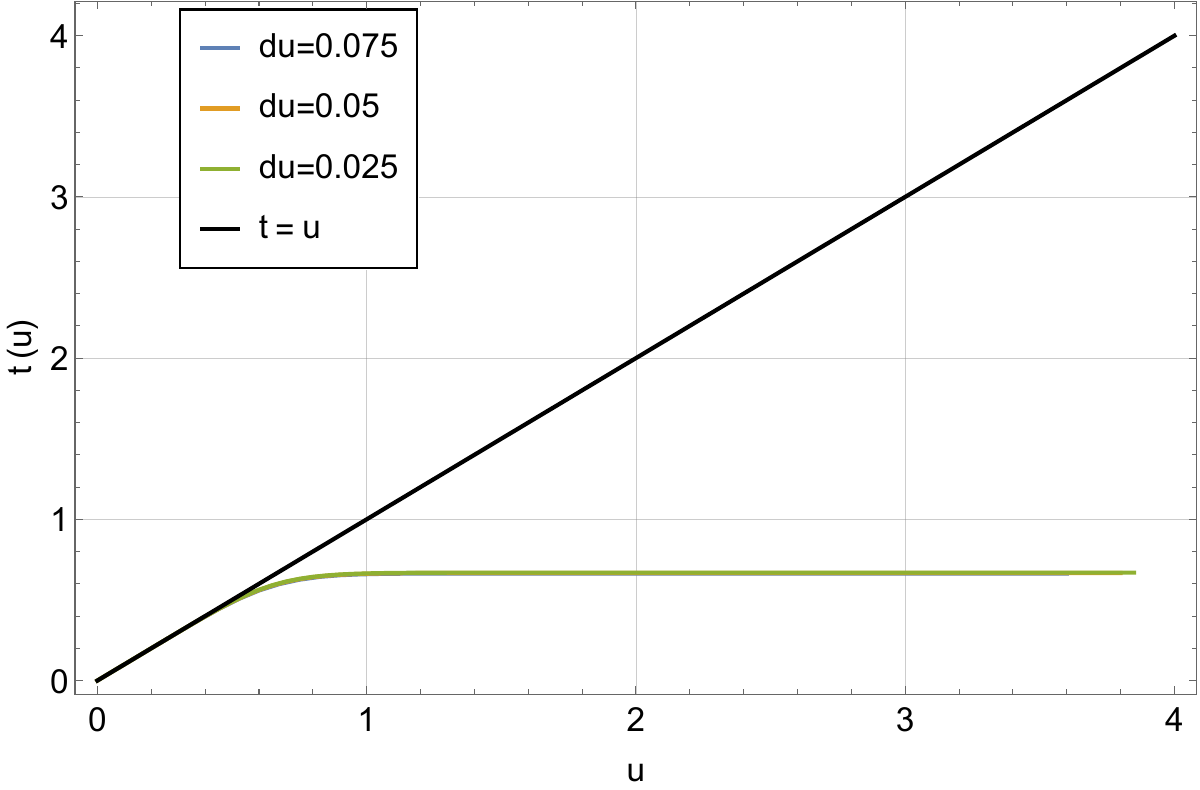}
      \end{minipage} &
      \begin{minipage}[t]{0.45\hsize}
        \centering
        \subcaption{$\dot{t}(u)$}
        \includegraphics[keepaspectratio, scale=0.35]{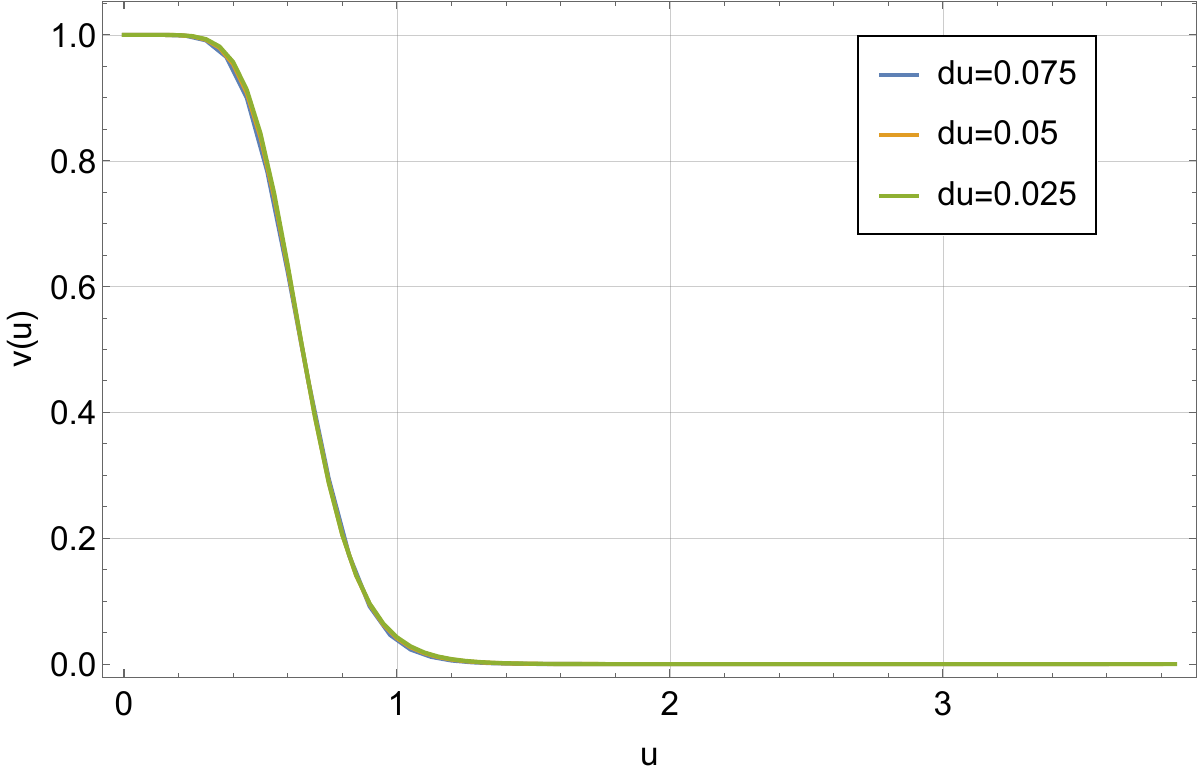}
      \end{minipage} \\[40pt]
   
      \begin{minipage}[t]{0.45\hsize}
        \centering
        \subcaption{$\log(\dot{t}(u))$}
        \includegraphics[keepaspectratio, scale=0.35]{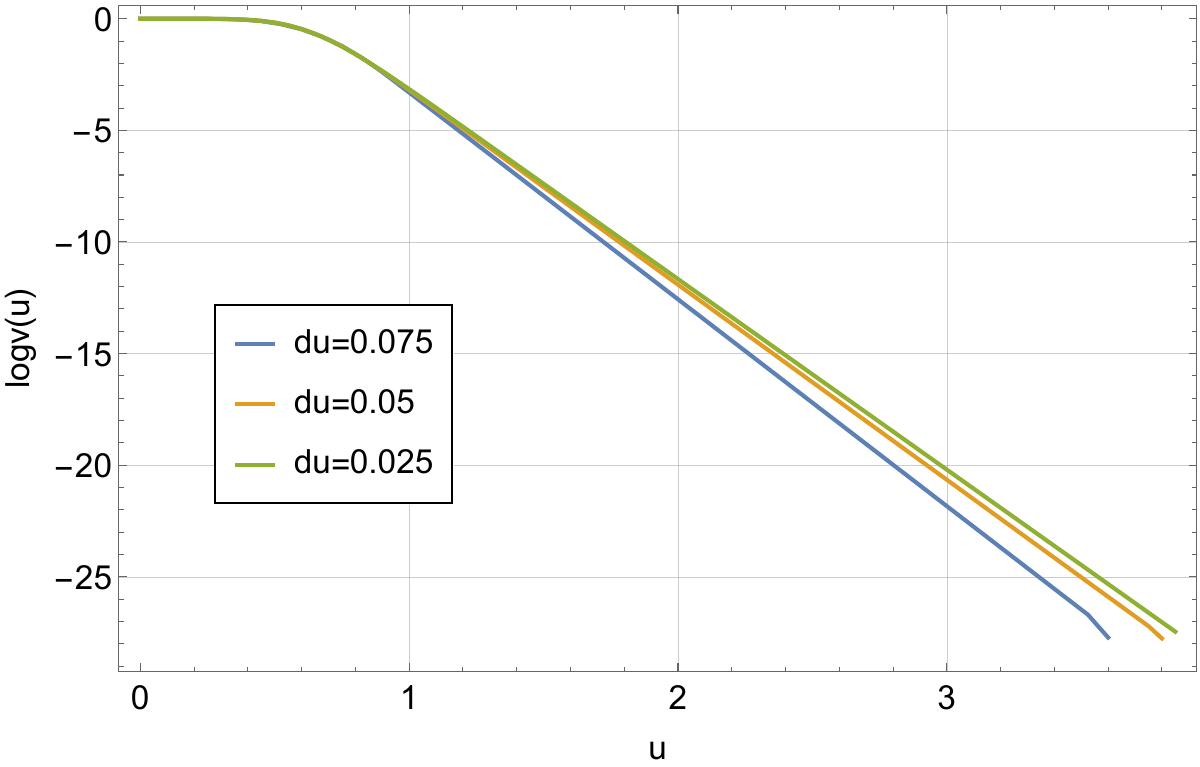}
      \end{minipage} &
      \begin{minipage}[t]{0.45\hsize}
        \centering
        \subcaption{$\{t,u\}$}
        \includegraphics[keepaspectratio, scale=0.35]{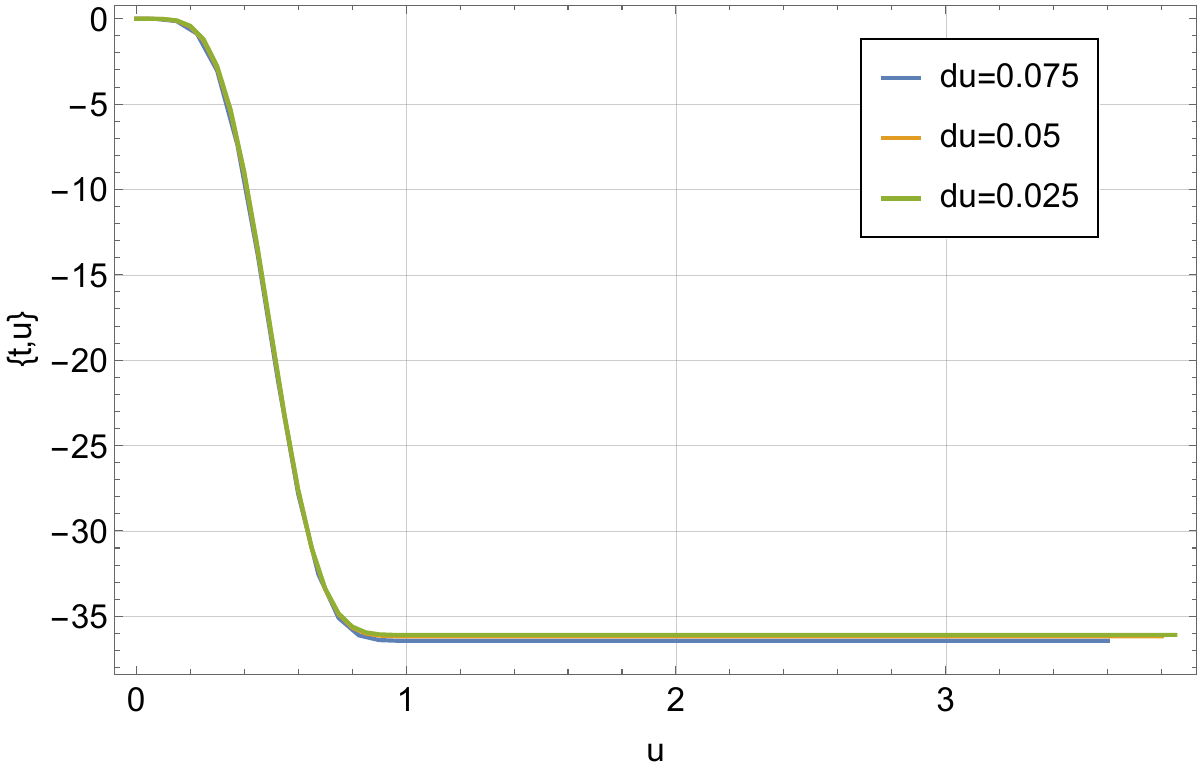}
      \end{minipage} 
    \end{tabular}
    \caption{Numerical convergence of our method with $\epsilon=0.1$ and $g(u)$ given in \eqref{g_cubic}. We see the convergence as step size $\mathrm{d}u$ becomes smaller.}
    \label{fig_append_D}
\end{figure}

In the main text, we have solved \eqref{eom} numerically in the following way.
We first define
\begin{align}
    v(u)=\dot t(u),\qquad a(u)=\ddot t(u),\qquad S(u)= \{t,u\}.
\end{align}
Then, after setting \(8\pi G=1\) and absorbing \((4\Delta-2)\sin(\pi\Delta)C_\Delta\) into the normalization of the coupling, \eqref{eom} is reduced to
\begin{align}
    \dot S(u) = v(t)\,\mathrm{Pf}\int_0^u \mathrm{d}u' \,\mathcal{A}(u,u').
\end{align}

For the numerical evaluation of the finite part, we do not discretize the symbol \(\mathrm{Pf}\) directly.  Instead, we subtract the required local singular terms of the integrand near \(u^\prime=u\), and add back their finite endpoint contribution.  For the present equation, the required local expansion has the form
\begin{align}
    \mathcal{A}(u,u^\prime)
    =
    \frac{\mathcal{A}_{2\Delta+1}(u)}{(u-u^\prime)^{2\Delta+1}}
    +
    \frac{\mathcal{A}_{2\Delta}(u)}{(u-u^\prime)^{2\Delta}}
    +
    \frac{\mathcal{A}_{2\Delta-1}(u)}{(u-u^\prime)^{2\Delta-1}}
    +
    \cdots.
\end{align}
Here the coefficients \(\mathcal{A}_{2\Delta+1}(u)\), \(\mathcal{A}_{2\Delta}(u)\), and \(\mathcal{A}_{2\Delta-1}(u)\) are determined by the local expansion of \(\mathcal{A}(u,u^\prime)\) itself.
Note that the final term is integrable, but included for numerical stability.
We then find
\begin{align}
    \mathrm{Pf}\int_0^u \d u^\prime\,\mathcal{A}(u,u^\prime)
    &=
    \int_0^u \d u^\prime\,
    \Bigg[
        \mathcal{A}(u,u^\prime)
        -
        \frac{\mathcal{A}_{2\Delta+1}(u)}{(u-u^\prime)^{2\Delta+1}}
        -
        \frac{\mathcal{A}_{2\Delta}(u)}{(u-u^\prime)^{2\Delta}}
        -
        \frac{\mathcal{A}_{2\Delta-1}(u)}{(u-u^\prime)^{2\Delta-1}}
    \Bigg]
    \notag\\
    &\quad
    -
    \frac{\mathcal{A}_{2\Delta+1}(u)}{2\Delta\,u^{2\Delta}}
    -
    \frac{\mathcal{A}_{2\Delta}(u)}{(2\Delta-1)u^{2\Delta-1}}
    -
    \frac{\mathcal{A}_{2\Delta-1}(u)}{(2\Delta-2)u^{2\Delta-2}} .
\end{align}
After this subtraction, the integrand in the first line is finite at \(u^\prime=u\), and hence it can be evaluated by an ordinary trapezoidal rule.
In addition, since the choices of $g(u)$ in section \ref{sec:Numerical verification} all behave $g(u) \sim u^3$ near $u=0$, the above surface terms do not diverge at $u=0$.

The time evolution is performed using a Heun-type predictor-corrector method.
In the implementation, we use one Euler predictor and then apply the trapezoidal corrector twice at each time step.
The convergence with respect to the step size is shown in Fig.\ \ref{fig_append_D}.

\bibliographystyle{jhep} 
\bibliography{ref.bib}

\end{document}